\documentclass[reprint,twocolumn,showpacs,fleqn,a4paper,floatfix,pra,aps,10pt]{revtex4-1}
\usepackage[T1]{fontenc}
\usepackage[latin1]{inputenc}
\ifx\pdftexversion\undefined
\usepackage[dvips]{graphicx}
\else
\usepackage[pdftex]{graphicx}
\fi
\usepackage{amsmath}% better  math
\usepackage{bm}% bold math
\usepackage{amssymb}% math symbols like Fraktur (\mathfrak{}) and Blackboard (\mathbb{})
\usepackage{latexsym}
\usepackage{pxfonts} %for lambdabar
\usepackage{dcolumn}
\usepackage{url}
\usepackage{tabularx}
\usepackage{multirow}
\usepackage[mediumqspace,mediumspace,amssymb,Gray]{SIunits}
\newcolumntype{d}{D{.}{.}{-1}}
\newcolumntype{f}[1]{D{.}{.}{#1}}
\newcommand{\greeksym}[1]{{\usefont{U}{psy}{m}{n}#1}}
\newcommand{\rmssmu}{{\mbox{\scriptsize{\greeksym{m}}}}}

\newcommand{\rd}{{\rm d}}
\newcommand{\re}{{\rm e}}
\newcommand{\ri}{{\rm i}}
\newcommand{\rD}{{\rm D}}
\newcommand{\rE}{{\rm E}}
\newcommand{\rN}{{\rm N}}
\newcommand{\rsp}{_{\rm p}}
\begin{document}

\title{Non-perturbative evaluation of some QED contributions to the muonic hydrogen  $\bm{n=2}$ Lamb shift and hyperfine structure}

\author{P.\ Indelicato}
\email{paul.indelicato@spectro.jussieu.fr}
\affiliation{
Laboratoire Kastler Brossel, École Normale Sup\' erieure; CNRS; Université Pierre et Marie Curie - Paris 6; 4, place Jussieu, 75252 Paris CEDEX 05, France
}

%\date{Received: \today / Revised version: date}
\date{November 1st, 2012}
%\pacs{31.30.jr}{QED corrections (Lamb shift) in muonic hydrogen and deuterium}
%\pacs{36.10.Ee}{Muonium, muonic atoms and molecules}
%\pacs{31.30.Gs}{Hyperfine interactions and isotope effects}

 % end of PACS codes 

\begin{abstract}
The largest contributions to the $n=2$ Lamb-shift, fine
structure interval and $2s$ hyperfine structure of muonic hydrogen are calculated by exact
 numerical evaluations of the Dirac equation, rather than by a perturbation
expansion in powers of $1/c$, in the framework of non-relativistic quantum electrodynamics.
Previous calculations and the validity of the perturbation expansion for
light elements are confirmed.   The dependence of the
various effects on the nuclear size and model are studied.
\end{abstract}

\maketitle

Despite many years of study, the proton charge radius has remained relatively
poorly known. It has been derived from measurements in electron-proton
collisions \cite{hmw1963,ssbw1980}  or from high-precision spectroscopy of hydrogen
\cite{npbj1992,whsl1995,bbnp1996,hugr1998,bnjc1997,dsaj2000,nhrp2000,pmah2011} as
described in the CODATA report in \cite{mtn2008}.  Tests of fundamental
physics based on the progress in accuracy of spectroscopy of hydrogen
and deuterium have been limited by the lack of an accurate value for the
proton radius.   Moreover, the
values for the proton radius obtained by different methods or different
analyses of existing experiments are spread over a range larger than the
uncertainty quoted for the individual results. Two
recent measurements have resulted in a puzzle.  The accurate determination
of the $2S$ Lamb shift by laser spectroscopy in muonic hydrogen provides
a proton size with a ten times smaller uncertainty than any previous
value and it differs by \emph{five} standards deviations from the 2006 CODATA
value\cite{pana2010}. At the same time,
a new, improved determination of the charge radius by electron
scattering, performed at Mainz with the MAMI microtron, provides a value
in good agreement with the value from hydrogen and deuterium
spectroscopy \cite{ber2010a,baab2010}.  Taking into account improved theory in hydrogen and deuterium and the MAMI measurement 
 lead the recently released 2010 adjustment \cite{CODATA2010} to differ by  6.9 standard deviation between from the proton radius obtained from muonic hydrogen. 

Many papers have been published in the last year, trying to solve this puzzle. A few are dealing with the calculation of the $n=2$ level energies in muonic hydrogen. Several others are concerned
 with the effect of the internal structure of the proton on these energies \cite{ruj2010,ruj2011,cav2011,cng2011,cam2011,car2011,dbw2011,mtcr2011,sic2011,vaw2011,hap2011a,sic2012}.
Others look at exotic phenomenon beyond the standard model \cite{jar2010,bckm2011,bmp2011,bab2011,rcg2011,tay2011,bckm2012}. 

Many contributions to the Lamb shift, fine, and hyperfine structure of
muonic hydrogen have been evaluated over the years; the results are
summarized in \cite{pac1996,vap2004,pac1999,egs2001,bor2005a,mar2005}
and in a recent book by Eides et al. \cite{egs2007}.  Most of these
calculations are done in the framework of nonrelativistic QED. The
wavefunction and operators are expanded in powers of the fine-structure
constant, and the contributions are obtained by perturbation theory.
Hylton \cite{hyl1985} showed that the
perturbation calculation of the finite size correction to the vacuum
polarization in heavy elements gives incorrect results. Since a bound
muon is closer to the nucleus than a bound electron by a factor
$m_\rmssmu/m_\re \approx 207$, its Bohr radius is slightly smaller than
the Compton wavelength of the electron $\lambdabar_{\rm C}=\hbar/m_\re
c$ by a factor  $m_\re/\alpha \, m_\rmssmu\approx 137/207$
($\alpha\approx 1/137.036$ is the fine structure constant, $m_\re$ and
$m_\rmssmu$ the electron and muon mass respectively).  The Compton
wavelength is the scale of QED corrections, and for the 2S level, the
muon wavefunction mean radius is only 2.6 times larger than the electron
Compton wavelength.

It is thus worthwhile to reconsider the largest corrections that
contribute to the 2S Lamb-shift in muonic hydrogen using
non-perturbative methods.  In the present work, we use the latest
version of the MCDF code of Desclaux and Indelicato \cite{iad2005},
which is designed to calculate properties of exotic atoms
\cite{spbi2005},  to evaluate the exact contribution of the electron
Uehling potential with Dirac wavefunctions including the finite nuclear
size.  In the same way, we calculate the Källén and Sabry contribution.

Throughout this paper we will use QED units, $\hbar=1$, $c=1$. The electric charge
is given by $e^2= 4\pi\alpha$.

\section{Numerical evaluation of the Dirac equation with realistic nuclear charge distribution models}
\label{sec:num-ev}

\subsection{Evaluation by the numerical solution of the Dirac equation}
\label{subsec:dirac}
We calculate higher-order  finite size correction, starting from the Dirac equation with reduced mass, as techniques for the accurate numerical solution of the Dirac equation in a Coulomb
potential have been developed over a period of many years within theframework of the Multiconfiguration Dirac-Fock (MCDF) method for the
atomic many-body problem \cite{gra1965,gra1970,dmo1971,des1975}.

The Dirac equation is written as
\begin{equation}
\label{eq:dirac}
\left[\boldsymbol{\alpha}\cdot\boldsymbol{p}+\beta \mu_{\mathrm{r}}
+V_{\mathrm{N}} (\boldsymbol{r}) \right] \Phi_{n \kappa \mu}
(\boldsymbol{r})   = \mathcal{E}_{n \kappa \mu}  \Phi_{n \kappa \mu}
(\boldsymbol{r}) ,
\end{equation}
where $\boldsymbol{\alpha}$ and $\beta$ are the Dirac $4\times4$
matrices, $V_{\mathrm{N}} (\boldsymbol{r})$ is the Coulomb potential of the nucleus, $\mathcal {E}_{n \kappa \mu} $
is the atom total energy, and $\Phi$ is a one-electron Dirac
four-component spinor:
\begin{equation}
\label{eq:diracspin}
    \Phi_{n \kappa \mu} (\boldsymbol{r}) = \frac{1}{r} \left[
    \begin{array}{c} P_{n \kappa}(r) \, \chi_{\kappa \mu}(\theta , \phi)
    \\[5 pt] \ri\, Q_{n \kappa}(r) \, \chi_{-\kappa \mu}(\theta , \phi)
    \end{array} \right]
\end{equation}
in which $\chi_{\kappa \mu}(\theta , \phi)$ is the two-component Pauli
spherical spinor \cite{gra1970}, $n$ is the principal quantum number,
$\kappa$ is the Dirac quantum number, and $\mu$ is the eigenvalue of
$J_z$.  This reduces, for a spherically symmetric potential, to the
differential equation:
%\begin{equation}
%\label{eq:dirac-diff}
%\left[
%\begin{array}{ccc}
%\alpha V_{\mathrm{N}} (r)&     -\frac{\rd}{\rd r} +\frac{\kappa}{r} \\[5 pt]
%\frac{\rd}{\rd r} +\frac{\kappa}{r} & \ \alpha V_{\mathrm{N}} (r) -2 \mu_{\mathrm{r}} c
%\end{array}
%\right]
%\left[
%\begin{array}{c}
%P_{n \kappa}(r)  \\[5 pt] 
%Q_{n \kappa}(r) 
%\end{array} 
%\right]
%= \alpha E^\rD_{n \kappa \mu} 
%\left[
%\begin{array}{c}
%P_{n \kappa}(r)  \\[5 pt]
%Q_{n \kappa}(r) 
%\end{array} 
%\right] ,
%\end{equation}
\begin{equation}
\label{eq:dirac-diff}
\left[
\begin{array}{ccc}
V_{\mathrm{N}} (r)&     -\frac{\rd}{\rd r} +\frac{\kappa}{r} \\[5 pt]
\frac{\rd}{\rd r} +\frac{\kappa}{r} & V_{\mathrm{N}} (r) -2 \mu_{\mathrm{r}}
\end{array}
\right]
\left[
\begin{array}{c}
P_{n \kappa}(r)  \\[5 pt] 
Q_{n \kappa}(r) 
\end{array} 
\right]
= E^\rD_{n \kappa \mu} 
\left[
\begin{array}{c}
P_{n \kappa}(r)  \\[5 pt]
Q_{n \kappa}(r) 
\end{array} 
\right] ,
\end{equation}
where $P_{n \kappa}(r)$ and $Q_{n \kappa}(r)$ are the large and small
radial components of the wavefunction, respectively, $\kappa$ the Dirac quantum number, $ E_{n \kappa
\mu}^\rD $ is the binding energy, $\mu_{\mathrm{r}}$ is the muon
reduced mass, $\mu_{\mathrm{r}}=m_{\rmssmu}M\rsp/(m_{\rmssmu}+M\rsp)$
($m_{\rmssmu}$ and $M\rsp$ are the muon and proton masses).

To solve this equation numerically, we use a 5 point predictor-corrector
method (order $h^7$) \cite{des1975,ddei2003} on a linear mesh defined as
\begin{equation}
\label{eq:mesh}
t_n = \ln\left(\frac{r_n}{r_0}\right) + a r_n,
\end{equation}
with $t_n=t_0 + nh$, and $r_0>0$ is the first point of the mesh,
corresponding to $n=0$. This immediately gives $t_0 = a r_0$. 
Equation~(\ref{eq:mesh}) can be inverted to yield
\begin{eqnarray}
r_n &= &\frac{W \left(ar_0  \mathrm{e}^{\,t_n}\right)}{a}, 
\nonumber \\
\frac{\rd r_n} {\rd t_n}&=&\frac{W \left(a r_0 \mathrm{e}^{\,t_n}\right)}
{a \left[1+W \left(a r_0 \mathrm{e}^{\,t_n)}\right)\right]},
\end{eqnarray}
where $W$ is the Lambert (or product logarithm) function. The
wavefunction and differential equation between 0 and $r_0$ are
represented by a 10 term series expansion.  For a point nucleus, the
first point is usually given by $r_0 = 10^{-2}/Z$ and $h=0.025$. Here we
use values down to $r_0 = 10^{-7}/Z$ and $h=0.002$ to obtain the best
possible accuracy. For a finite charge distribution, the nuclear
boundary is fixed at the value $r_N$, where $N$ is large enough to
obtain sufficient accuracy.   The mean value of an operator $\mathcal{O}$, that gives the
first-order contributions to the energy, is calculated as
\begin{eqnarray}
\Delta E_{\mathcal{ O}}&=&\int_0^{\infty} \rd r \left[P(r)^2+Q(r)^2\right] \mathcal{O}(r)\nonumber \\
&=& \int_0^{r_0} \rd r \left[P(r)^2+Q(r)^2\right]  \mathcal{O}(r)\nonumber \\
&&+\int_{r_0}^{\infty} \rd t \frac{\rd r}{\rd t}\left[P(r)^2+Q(r)^2\right] \mathcal{O}(r)
\end{eqnarray}
using 8 and 14 points integration formulas due to Roothan. The two
integration formulas provide the same result within 9 decimal places.

\subsection{Charge distribution models}
\label{subsec:charmod}

For the proton charge distribution, two models are extensively used.
The first corresponds to a proton dipole (charge) form factor, the
second is a gaussian model. Here we also use uniform and Fermi charge
distributiond and fits to experimental data \cite{bhm2007,amt2007}. The
analytic distributions are parametrized so they provide the same mean
square radius $R$. Moments of the charge distribution are defined by
\begin{equation}
\label{eq:rmom}
<r^n>=4 \pi \int_0^{\infty} r^{2+n} \rho(r) \rd r,
\end{equation}
where the nuclear charge distribution $\rho(r) = \rho_{\rN}(r)/(Z e)$ is
normalized by
\begin{equation}
\label{eq:norm-rho}
	 \int \rho(\vec{r})  \rd \vec{r} =4 \pi \int_{0}^{\infty} \rho(r) r^{2} \rd r = 1, 
\end{equation}
for a spherically symmetric charge distribution.
The mean square radius is $R=\sqrt{<r^2>}$.

The potential can be deduced from the charge density using the well
known expression:
\begin{equation}
\label{eq:poisson}
V_{\mathrm{N}} (r) = - \frac{4\pi e}{r} \int_0^r \rd u \, u^2 \rho_\rN(u) 
- 4\pi e\int_r^{\infty} \rd u u \rho_\rN(u) .
\end{equation}

The exponential charge distribution and corresponding potential energy are written 
\begin{eqnarray}
\label{eq:expo}
\rho_\rN(r) &=& Z e\,\frac{\mathrm{e}^{-\frac{r}{\xi}}}{8\pi \xi^3}, \nonumber
\\[4 pt]
V_{\mathrm{N}} (r) &=& -Ze^2\left(\frac{1-\mathrm{e}^{-\frac{r}{\xi}}}{r} - 
\frac{\mathrm{e}^{-\frac{r}{c}}}{2\xi}\right), \nonumber \\[4 pt]
<r^n>&=&\frac{(n+2)!\,\xi^n}{2},
\end{eqnarray} 
which gives $\xi=\frac{R}{2\sqrt{3}}$.
The gaussian charge distribution and potential are given by
\begin{eqnarray}
\label{eq:gauss}
\rho_\rN(r) &=& Ze\,\frac{\mathrm{e}^{-\left(\frac{r}{\xi}\right)^2}}
{\pi^{3/2}\xi^3}, \nonumber \\
V_{\mathrm{N}} (r) &=& -Ze^2\,\frac{\textrm{erf}\left(\frac{r}{\xi}\right)}{r}, 
\nonumber \\
<r^n>&=&\frac{2 \Gamma\left(\frac{n+3}{2}\right)\xi^n}{\sqrt{\pi}},
\end{eqnarray} 
where $\mathrm{erf}$ is the error function, and $\xi=\sqrt{\frac{2}{3}}R$. Other similar expressions for the two above models can be found in \cite{fri1979}.

\begin{figure}[htbp]
\begin{center}
\includegraphics[width=0.5\columnwidth,trim =11cm 10cm 4cm 13cm,clip]{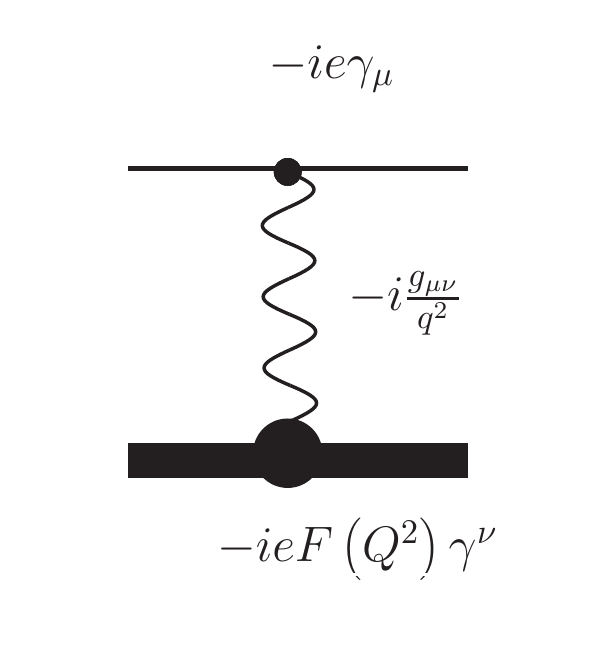}
\caption{Electromagnetic interaction between a lepton (narrow line) and a nucleon (bold line).}
\label{fig:elec-nucl-int}
\end{center}
\end{figure}

The electric form factor is related to the charge distribution by
\begin{equation}
\label{eq:formfac}
G_\rE (\vec{Q}^2)=\int \rd \bm r e^{-\ri \boldsymbol{q}\cdot\boldsymbol{r}} 
%\rho_\rN(\boldsymbol{r}),
% so that G(0) is 1, otherwise should be Z e
\rho(\boldsymbol{r}),
\end{equation}

For the exponential model this leads to
\begin{equation}
\label{eq:exp-formfac}
G_\rE (\vec{Q}^2)=\frac{1}{\left(1+\frac{R^2 \bm q^2}{12}\right)^2}\approx 1 - 
\frac{R^2}{6}\bm q^2+\frac{R^4}{48}\bm q^4+ \cdots
\end{equation}
while for the Gaussian model one has
\begin{equation}
\label{eq:gauss-formfac}
G_\rE (\vec{Q}^2)=e^{-\frac{1}{6}R^2 \bm q^2} \approx 1 -
\frac{R^2}{6}\bm q^2+\frac{R^4}{72}\bm q^4+ \cdots
\end{equation}
The two models have an identical slope $R^2/6$ as functions of $\bm q^2$
for $\bm q\to0$ as expected (see, e.g, \cite{pil2005}).

In 1956, Zemach introduced an electromagnetic form factor, useful for
evaluating the hyperfine structure energy correction
\begin{equation}
  \label{eq:em-form-factor}
  \rho_{em} (\boldsymbol{r}) = \int \rho(\boldsymbol{r}-\boldsymbol{u})\mu(\boldsymbol{u})d\boldsymbol{u} ,
\end{equation}
where $\mu(\boldsymbol{u})$ is the magnetic moment density. Both $\mu(\boldsymbol{u})$ and  $ \rho_{em}$ are normalized to unity as in Eq. \eqref{eq:norm-rho}. 
The Zemach radius is given by
\begin{equation}
  \label{eq:zem-rad}
R_{\textbf{Z}}=  \left<r_{\textbf{Z}}\right> = \int r \rho_{em} (\boldsymbol{r}) d\boldsymbol{r}.
\end{equation}
The Zemach's radius can be written in momentum space as \cite{bchh2005,kar1999}
\begin{equation}
\label{eq:zemq}
R_{\textbf{Z}}=\frac{-4}{\pi} \int dq \frac{1}{q^2}\left(G_\rE \left(\vec{Q}^2\right) \frac{G_{\textrm{M}}\left(\vec{Q}^2\right)}{1+\kappa_p}-1\right),
\end{equation}
where $\kappa_p$ is the proton anomalous magnetic moment, and $G_{\textrm{M}}$ is normalized so that $G_{\textrm{M}}\left(0\right)=1+\kappa_p$.
The exponential and gaussian models enables to obtain analytic results for  $R_{Z}$ as a function of the charge and magnetic moment radii $R$ and $R_{\textrm{M}}$.
Using \eqref{eq:exp-formfac} or \eqref{eq:gauss-formfac} for the exponential or gaussian model, and Eq. \eqref{eq:zemq}, we get respectively 
\begin{equation}
\label{eq:zemrad-exp}
R^{\textrm{Exp.}}_{\textbf{Z}}=\frac{3 R^4+9 R^3 R_{\textrm{M}}+11 R^2 R_{\textrm{M}}^2+9 R R_{\textrm{M}}^3+3R_{\textrm{M}}^4}{2 \sqrt{3} (R+R_{\textrm{M}})^3}
\end{equation}
\begin{equation}
\label{eq:zemrad-gauss}
R^{\textrm{Gauss}}_{\textbf{Z}}=2 \sqrt{\frac{2}{3 \pi }} \sqrt{R^2+R_{\textrm{M}}^2}.
\end{equation}

An other useful quantity, which appears in the estimation of the finite size correction to vacuum polarization is the third Zemach's moment
\begin{equation}
  \label{eq:zem-rad3}
 \left<r^3\right>_{(2)} = \int r^3 \rho_{(2)} (\boldsymbol{r}) d\boldsymbol{r}, 
\end{equation}
where the  convolved charge distribution is
\begin{equation}
  \label{eq:e2-form-factor}
  \rho_{(2)} (\boldsymbol{r}) = \int \rho(\boldsymbol{r}-\boldsymbol{u})\rho(\boldsymbol{u})d\boldsymbol{u}.
\end{equation}
This can be rewritten in the more convenient form  \cite{pac1996,ruj2011}, in the limit of large proton masses, 
\begin{equation}
  \label{eq:zem-rad3q}
 \left<r^3\right>_{(2)} = \frac{48}{\pi} \int dq \frac{1}{q^4}\left( G^2_{\rE} \left(\vec{Q}^2\right)-1+\frac{q^2}{3}R^2\right).
 \end{equation}
It can be easily seen from Eqs. \eqref{eq:exp-formfac} or  \eqref{eq:gauss-formfac} that the expression is finite for $q\to 0$.

We now turn to more realistic models, based on experiment. A recent analysis of the world's data on elastic electron-proton scattering and calculations of two-photon exchange effects provides an analytic expression for the electric  form factors \cite{amt2007}, given as
\begin{equation}
\label{eq:form-fac-exper}
G_{\rE}\left(\vec{Q}^2\right)=\frac{1+\sum_{i=0}^2 a_i \tau^i}{1+\sum_{j=0}^4 b_j \tau^j},
\end{equation}
where  $\tau=q^2/(4 M_p)$. The $a_i$  coefficients can be found in Table I of Ref. \cite{amt2007}. A second work \cite{bhm2007} uses a combination of several spectral functions tacking into account several resonances and continua like the $2\pi$, $K\bar K$and $\rho \pi$ continua.  Here we use the fit resulting from the \emph{superconvergence approach} from this work. This corresponds to a sum of 12 dipole-like functions, which are able to represent the experimental data with a reduced $\chi^2$ of 1.8. A comparison of the electric form factor from both works is presented on Fig. \ref{fig:aring-vs-belu}. It is clear that both experimental form factors and the dipole approximation with an identical radius are very close. We can obtain the charge radius and the next correction by performing an expansion in $q$ of the experimental form factors. 
We get for Ref. \cite{amt2007}
\begin{equation}
\label{eq:exp-formfac-arri}
G_\rE \left(\vec{Q}^2\right) \approx 1 - \frac{0.8503^2}{6}q^2+\frac{0.8503^4}{43.3909}q^4+ \cdots \, ,
\end{equation}
and for Ref. \cite{bhm2007}
\begin{equation}
\label{eq:exp-formfac-belu}
G_\rE\left(\vec{Q}^2\right) \approx 1 - \frac{0.84995^2}{6}q^2+\frac{0.84995^4}{38.793}q^4+ \cdots \, .
\end{equation}
These expansions are very close to the one for a dipole form factors from Eq. \eqref{eq:exp-formfac}.
 
\begin{figure}[htbp]
\begin{center}
\includegraphics[width=\columnwidth]{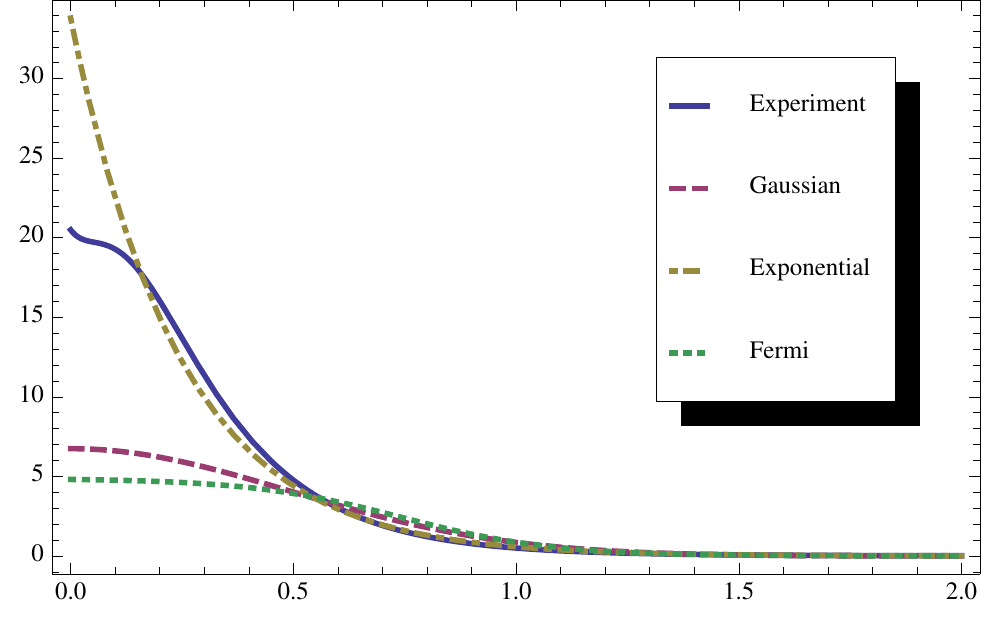}
\includegraphics[width=\columnwidth]{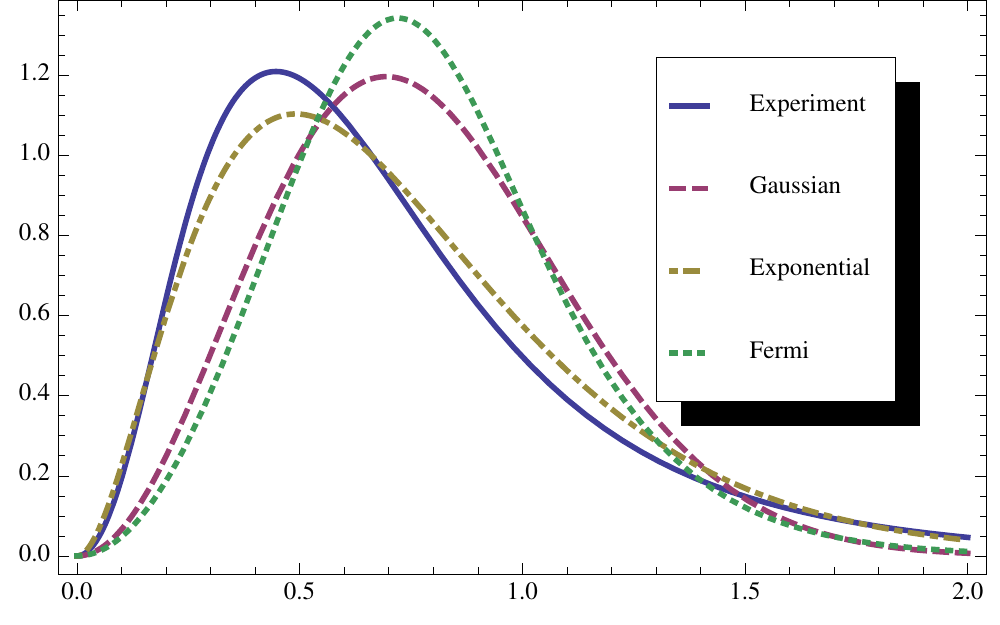}
\caption{Top: charge densities $\rho(r)$, bottom: charge densities $r^2 \rho(r)$  for the experimental fits in Ref. \cite{amt2007}, compared to Gaussian, Fermi and exponential models distributions. All models are calculated to have the same \unit{R=0.850}{fm} RMS radius as deduced from the experimental function.}
\label{fig:charge-dens}
\end{center}
\end{figure}

\begin{figure}[htbp]
\begin{center}
\includegraphics[width=\columnwidth]{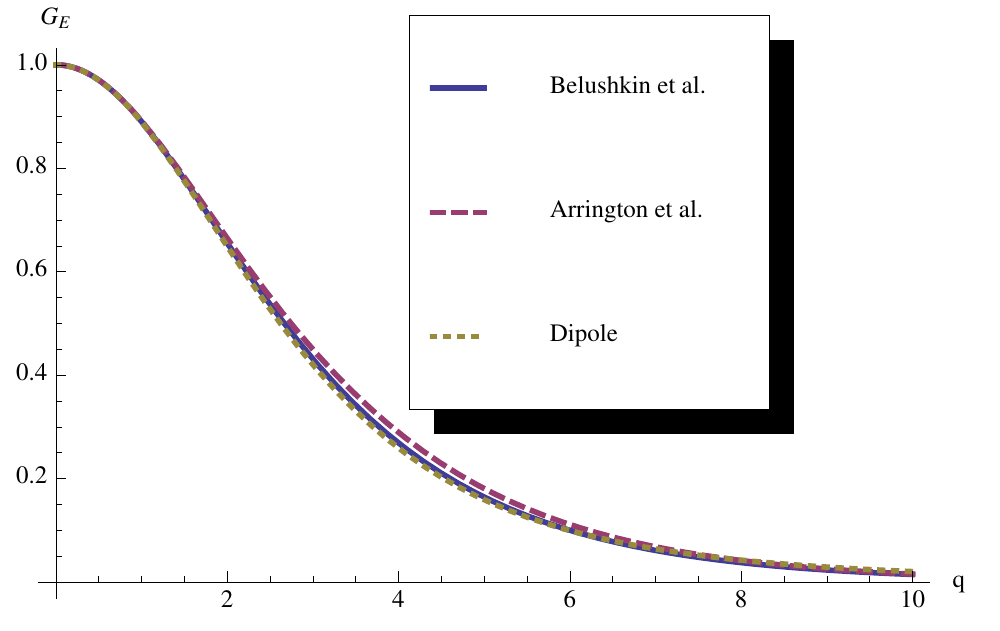}
\caption{Comparison of the electric form factor from Refs. \cite{amt2007,bhm2007}, with a dipole model with the same \unit{R=0.850}{fm} as deduced from the experimental function.}
\label{fig:aring-vs-belu}
\end{center}
\end{figure}

In order to compare different charge density  models, we have performed an analytic evaluation of the charge  densities corresponding to \cite{amt2007}, replacing  \eqref{eq:form-fac-exper} in \eqref{eq:formfac}  and performing the inverse Fourier transforms, to obtain the corresponding charge distribution, depending on the set of $a$  coefficients. 
The corresponding densities are plotted and compared to Fermi, Gaussian and exponential models. Distances are converted from GeV to fm using \unit{\hbar c=0.197326 9631}{GeVfm} in the density obtained from the experiment. 
The charge distributions are compared on Fig. \ref{fig:charge-dens}. We plotted both $\rho(r)$ and $\rho(r)r^2$ to reveal the differences at long and medium distances. The experimental charge density is rather different from all three analytic distributions, while, once multiplied by $r^2$, it is closer to the exponential distribution.

\section{Evaluation of main vacuum polarization and finite size correction}
\label{sec:vacpol}

\begin{figure}[tb]
	\centering
%====================
\includegraphics[width=\columnwidth]{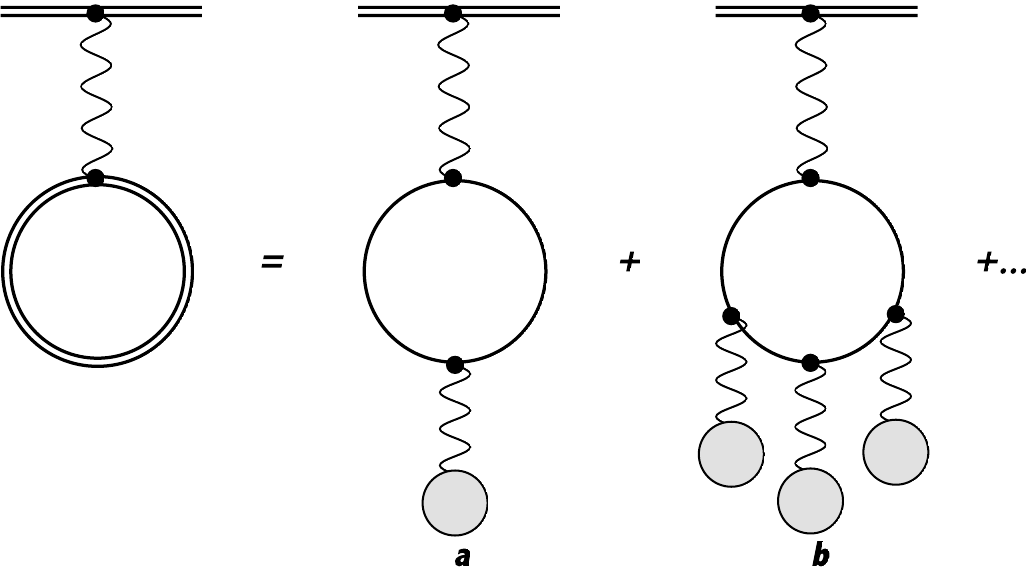}
	\caption[]{Feynman diagrams corresponding to  the full vacuum polarization contribution, and expansion in $Z\alpha$. Diagram (a) corresponds to the Uehling potential [Eqs. \eqref{eq:v11pn} and \eqref{eq:v11}]. Diagram (b) corresponds to the Wichmann and Kroll correction. The double line represents a 
	bound lepton wavefunction, the wavy line a 
	retarded photon propagator.  The single line correspond to a free electron-positron (or muon-antimuon) pair. The grey circles correspond to the interaction with the nucleus.} \label{fig:vp11}
\end{figure}

The Feynman diagram corresponding to the Uehling approximation to the vacuum polarization correction is presented in Fig. \ref{fig:vp11} (a). The evaluation of the vacuum polarization can be performed using  standard techniques of (perturbative) non-relativistic QED (NRQED) as described in \cite{pac1996,pac1999}. Here we use the analytic results of Klarsfeld \cite{kla1977}
as described in \cite{bai2000} and numerical solution of the Dirac equation from Sec. \ref{sec:num-ev}. In order to obtain higher order effects, we solve the Dirac equation in a combined potential resulting from the finite nuclear charge distribution and of the Uehling potential. 
The logarithmic singularity of the Uehling potential at the origin for a point charge cannot be easily incorporated in a numerical
Dirac solver. In the case of a finite charge distributions, the singularity is milder, but great care must be exercised to obtain results accurate enough for our purpose.

For a point charge, the Uehling potential, which represents the leading contribution to the vacuum polarization, is expressed as \cite{kla1977,ueh1935,bbeg1968}
\begin{equation}
 \label{eq:v11pn}
 \begin{split}
 V_{11}^{\mathrm{pn}}(r)& =-\frac{\alpha(Z\alpha)}{3\pi}\int_1^{\infty} dz  \sqrt{z^2-1}  \left(\frac{2}{z^2}+\frac{1}{z^4}\right)\frac{e^{-2 m_e r z}}{r}  \\
 &=-\frac{2\alpha(Z\alpha^2)}{3\pi} \frac{1}{r}\chi_{1}\left (\frac{2}{\lambda_e} r  \right)  \,
 \end{split}
\end{equation}
where $m_e$ is the electron mass, $\lambda_{e}$ is the electron Compton wavelength and the function $\chi_{1}$ belongs to a family of functions defined by
\begin{equation}
	\chi_{n}(x) = \int_{1}^{\infty}  dz e^{-x z} \frac{1}{z^{n}} \left(
	\frac{1}{z} + \frac{1}{2 z^{3}} \right) \sqrt{z^{2}-1}.
	\label{eqn:chi}
\end{equation}
The Uehling potential for a spherically symmetric charge distribution is expressed  as \cite{kla1977}
\begin{eqnarray}
	V_{11}(r)&=& - \frac{2\alpha (Z\alpha)}{3} \frac{1}{r}
	\int_{0}^{\infty} dr'\, r' \rho(r') \nonumber \\
	&&\times \left[ \chi_{2} \left(\frac{2}{\lambda_e} \mid r - r'
	\mid \right) - \chi_{2} \left( \frac{2}{\lambda_e}  \mid r + r' \mid \right) \right] \,.
	\label{eq:v11}
\end{eqnarray}
The expression of the potential at the origin is given by
\begin{equation}
\label{eq:v11(0)}
	V_{11}(0) = - \frac{8 \alpha (Z \alpha)}{3} \int_{0}^{\infty} dr' r'
	\rho(r') \chi_{1}\left(\frac{2}{\lambda_e} r' \right), 
\end{equation}
while it behaves at large distances as \cite{far1976}
\begin{eqnarray}
	V_{11}(r) &=& - \frac{2 \alpha (Z \alpha)}{3 \pi} \frac{1}{r}
	\left[ \chi_{1}\left(\frac{2}{\lambda_e} r\right) + \frac{2}{3} <r^{2}> \chi_{-1}\left(\frac{2}{\lambda_e} r \right)\right. \nonumber \\
	&& \qquad+\left. 
	\frac{2}{15} <r^{4}> \chi_{-3}(\frac{2}{\lambda_e} r) + \ldots \right]
	\label{eq:v11asymp}
\end{eqnarray}
using  the moments of the charge distribution \eqref{eq:rmom}.
The energy shift associated with the potential \eqref{eq:v11pn} or \eqref{eq:v11} in first order perturbation is calculated as
\begin{equation}
\Delta E^{11,\mathrm{pn}}_{nl\kappa}=
\left\langle n, l, \kappa, \mu_{\mathrm{r}}
 \left| 
  V_{11}
 \right|
  n, l, \kappa,\mu_{\mathrm{r}} \right\rangle
\end{equation}
where $\left| n, l, \kappa,\mu_{\mathrm{r}} \right\rangle$ is a wavefunction solution of Eq. \eqref{eq:dirac}, which depends on the reduced mass. It was shown recently \cite{jen2011a} that this method
provides the correct inclusion of the vacuum polarization recoil correction at the Barker and Glover level \cite{bag1955}. Since we directly use relativistic functions, the other corrections described in  \cite{jen2011a} are automatically included.

\section{Higher order QED corrections}
\label{sec:HOQED}
\subsection{Reevaluation of the Källèn and Sabry potential}
\label{subsec:v21hp}
\begin{figure}[tb]
	\centering
%====================
\includegraphics[width=\columnwidth]{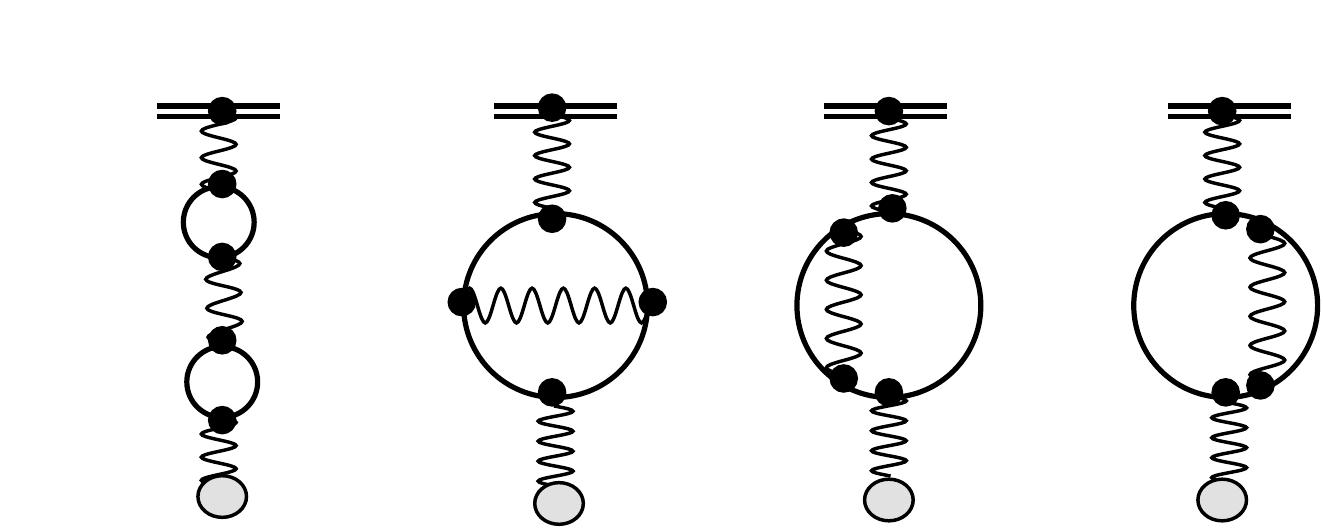}
	\caption[]{Feynman diagrams included in the Källen and Sabry $V_{21}(r)$ potential (Eq. \eqref{eq:kspot}). See Fig. \ref{fig:vp11} for explanation of symbols.} \label{fig:kas}
\end{figure}

The Källén and Sabry potential \cite{kas1955}, is a fourth order potential, corresponding to the diagrams in Fig. \ref{fig:kas}. The expression for this potential has also been derived on Ref. \cite{bmr1970,bmr1972a,bmr1972b,bar1973}. In the previous version of the \emph{mdfgme} code, the  Källèn and Sabry potential used was the one provided by Ref. \cite{far1976}, which is only accurate to 3 digits.
The expression of this potential is for a point charge
\begin{equation}
V_{21}(r) = \frac{\alpha^2(Z\alpha)}{\pi^2 r}L_{1}(\frac{2}{\lambda_e}r),
\label{eq:kspot}
\end{equation}
where
\begin{equation}
\label{eq:v21}
\begin{split}
L_{1}(r) = & \int_1^{\infty} dt e^{-r t} \Biggl(\left(\frac{2}{3 t^5}-\frac{8}{3 t}\right) f(t)  \\
& +\left(\frac{2}{3 t^4}+\frac{4}{3  t^2}\right) \sqrt{t^2-1} \ln \left(8 t \left(t^2-1\right)\right) \\ 
& +\sqrt{t^2-1}
   \left(\frac{2}{9 t^6}+\frac{7}{108 t^4}+\frac{13}{54 t^2}\right) \\
 & +\left(\frac{2}{9t^7}+\frac{5}{4 t^5}+\frac{2}{3 t^3}-\frac{44}{9 t}\right) \ln
   \left(\sqrt{t^2-1}+t\right)\Biggr),
   \end{split}
\end{equation}
and 
\begin{equation}
\label{eq:v21func}
f(t)=\int_t^{\infty} dx \left[\frac{\left(3 x^2-1\right) \ln \left(\sqrt{x^2-1}+x\right)}{x \left(x^2-1\right)}-\frac{\ln
   \left(8 x \left(x^2-1\right)\right)}{\sqrt{x^2-1}}\right].
\end{equation}
The function $f(t)$ can be calculated analytically in term of the $\ln$ and dilogarithm functions.
Blomqvist \cite{blo1972} has shown that $L_{1}(r)$ can be expressed as
\begin{equation}
\label{eq:L1shape}
L_{1}(r)=g_2(r)\ln^2(r)+g_1(r)log(r)+g_0(r),
\end{equation}
and provided a series expansion of this function for small $r$.
Fullerton and Rinker \cite{far1976} provided polynomial approximations to the functions $g_i(r)$. Here we have numerically evaluated the function $L_{1}(r)$ to a very good accuracy, using Mathematica. We then fitted the coefficients of polynomials for the function $g_i(r)$. The results are presented in Appendix \ref{app:v21pn}.
For $x>3$, we have used the functional form
\begin{equation}
\label{eq:l1asymp}
L_{1}(r)=\frac{e^{-r} \left(a+b \sqrt{r}+c r+d r^{3/2}+e r^2+f r^{5/2}\right)}{r^{7/2}}.
\end{equation}
The coefficients are also given in Appendix  \ref{app:v21pn}.

To obtain the finite nuclear size correction, we use the known expression for a spherically-symmetric charge distribution \cite{far1976}
\begin{eqnarray}
V_{21}(r) &=& \frac{\alpha^2(Z\alpha)}{\pi^2 r}\int_0^{\infty} dr' r' \rho(r')\biggl(L_{0}(\frac{2}{\lambda_e}\mid r-r'\mid) \nonumber \\
                 &&\qquad \qquad \qquad-L_{0}(\frac{2}{\lambda_e}\mid r+r'\mid)\biggr),
\end{eqnarray}
where
\begin{equation}
L_{0}(x) = -\int^x du L_1(u).
\end{equation}
Using our approximation to $L_1(x)$ in Eqs \eqref{eq:L1shape} and \eqref{eq:l1asymp}, we obtain the following approximate expressions for $L_0(x)$. 
For $x\leq 3$, the expression is very similar to the one for $L_1(x)$. One obtains
\begin{equation}
\label{eq:L0shape}
L_{0}(r)=r h_2(r)\ln^2(r)+ r h_1(r)log(r)+ h_0(r),
\end{equation}
The expression for the functions $h_i$ are given in Appendix \ref{app:v21}.
For the asymptotic function, given for $x>3$, we integrate directly Eq. \eqref{eq:l1asymp}, which yield (fixing the integration constant so that $L_0(r)$ is 0 at infinity)
\begin{equation}
\label{eq:l0asymp}
\begin{split}
L_0(r) & = 41.1352787432251923 \\ 
&-5.1094977559522696 \Biggl(8.05074798111 \text{erf}\left( \sqrt{r}\right) \\
&\qquad -1.028091975364 \text{Ei}(-r) \\
&+\frac{e^{-r}}{r^{5/2}}
   \biggl(-2.02809197536 r^{3/2}+4.54214815071 r^2 \\
   &  \qquad-0.494718704003 r+0.98439728916 \sqrt{r} \\
   & \qquad -0.344009752879\biggr)\Biggr)
      \end{split}
\end{equation}
where $\text{Ei}(r)$ is the exponential integral.

\section{Numerical Results}
\subsection{Finite size correction to the Coulomb contribution}
Obtaining the accuracy required from the calculation on $E^D_{2 \kappa \mu} $, which has a value of $\approx$ \unit{632.1}{eV}, while the Lamb shift is $\approx$\unit{0.22}{eV} with an aim at better than \unit{0.001}{meV}, is a very demanding task.
For a point nucleus, we get exact degeneracy for the $2s$ and $2p_{1/2}$ Dirac energies.  
The best numerical accuracy was obtained generating the wavefunction on a grid with  $r_0=2\times 10^{-3}$ and $h=2\times 10^{-3}$. This corresponds to $\approx 8700$ tabulation points for the wavefunction, with around 2800 points inside the proton.  I checked that variations in $r_0$ and $h$ do not change the final value. The main finite nuclear size effect on the $2p_{1/2}$--$2s$ energy 
separation comes from the sum of the Dirac energy splitting (the $2p_{1/2}$ and $2s$ level are exactly degenerate for a point nucleus).

I evaluated the different quantities on a grid of proton sizes ranging from 0.3 to \unit{1.2}{fm}, with steps of \unit{0.025}{fm} (80 points). I also evaluated the contribution for the muonic hydrogen proton size and the CODATA 2010 proton size. The first few terms of the dependence of the relativistic energy on the moments of the charge distribution where given by Friar \cite{fri1979}. For a Gaussian charge distribution he finds for a $s$ state
\begin{equation}
\begin{split}
\Delta E^{\mathrm{Coul.}}&=a \langle r^2 \rangle + b \langle r^3 \rangle _{(2)} +c \langle r^2 \rangle^2 \\
 & + d  \langle r^2 \rangle \langle \log r \rangle + e  \langle r^2 \rangle^2 \langle \log r \rangle.
 \end{split}
\end{equation}
I use this as a guide to fit my numerical results.

\begin{figure}[htbp]
\begin{center}
\includegraphics[width=\columnwidth]{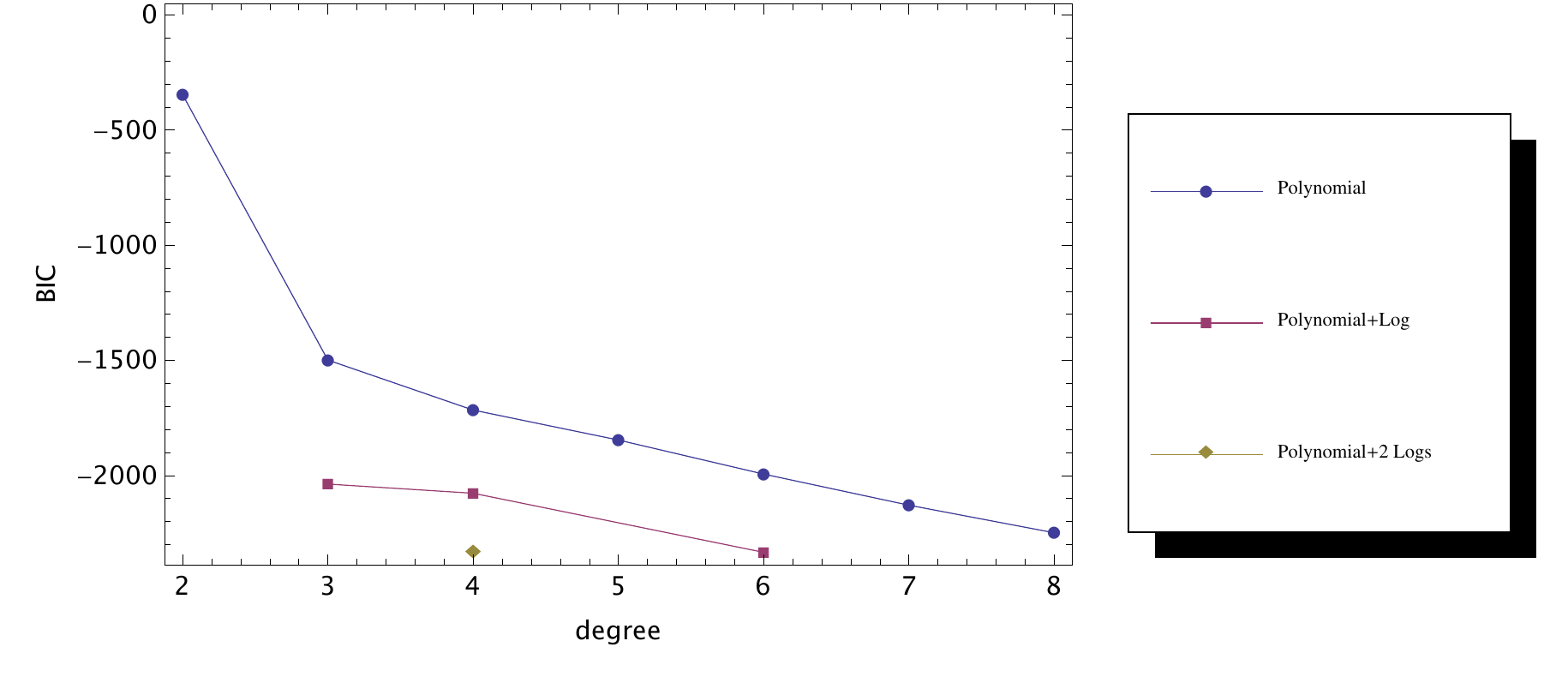}
\includegraphics[width=\columnwidth]{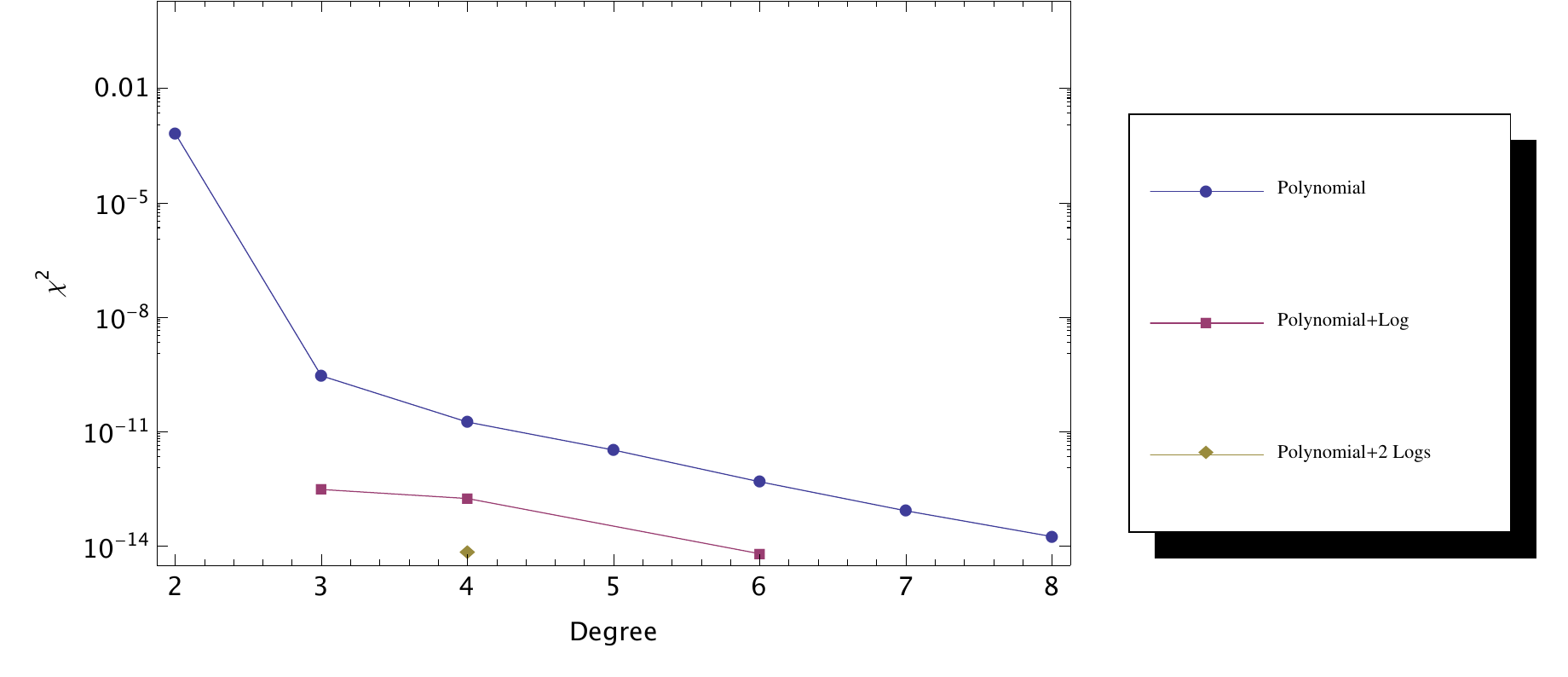}
\caption{BIC criterium (top) and reduced $\chi^2$ (bottom) as function of the degree of the polynomial and of the logarithmic dependence used in the fit of the energy.}
\label{fig:fit-qual}
\end{center}
\end{figure}

As a first example a 3-parameter fits provides
\begin{eqnarray}
\label{eq:coulfsr3}
\Delta E^{\mathrm{Coul.}} _{2p_{1/2}-2s_{1/2}}(R)&=& -5.19972\, R^2 +0.0351289\, R^3 \nonumber \\
 && -0.0000534235\, R^4 \, \mathrm{meV}
\end{eqnarray}
A better fit is provided by
\begin{equation}
 \begin{split}
\label{eq:coulfsr6}
\Delta E^{\mathrm{Coul.}} _{2p_{1/2}-2s_{1/2}}(R)&=-5.19990 R^2 +0.0355905 R^3   \\
 & -0.000488059 R^4+0.000172334 R^5 \\ 
 &-0.0000245051 R^6 \, \mathrm{meV}
\end{split}
\end{equation}
Using Friar functional form with only one $\log R$ term, I obtain
\begin{equation}
 \begin{split}
\label{eq:coulfsr6log}
\Delta E^{\mathrm{Coul.}} _{2p_{1/2}-2s_{1/2}}(R)&=-5.199365 R^2 + 0.03466100 R^3 \\
&+ 0.00007366037 R^4 \\
& -  0.00001720960 R^5 \\
 &+ 1.198332\times10^{-6} R^6 \\
& + 0.0002677236 R^2 \log R \, \mathrm{meV}.
\end{split}
\end{equation}
The function with two logarithmic terms and close values of the BIC and $\chi^2$ criteria is given by
\begin{equation}
 \begin{split}
\label{eq:coulfsr6log2}
% checked July 21st, 2012
\Delta E^{\mathrm{Coul.}} _{2p_{1/2}-2s_{1/2}}(R)&=-5.199337 R^2 \\
&+ 0.03458139 R^3 + 0.0001092856 R^4 \\
&+  0.0002788380 R^2 \log R \\
&- 0.00004957598 R^4 \log R
\end{split}
\end{equation}
Criteria for the quality of the fit are plotted in Fig.\ \ref{fig:fit-qual}. I use both the reduced $\chi^2$ and a Bayesian information criterion (BIC) to evaluate the improvement in the value when increasing the number of parameters \cite{baa2004}. 
 We obtain a coefficient for $R^2$ which is \unit{-5.1999}{meV/fm^2} and a coefficient for $R^3$ equal to  \unit{0.03559}{meV/fm^3}.  Using our numerical solutions we also find \unit{-5.19972}{meV/fm^2} and \unit{0.032908}{meV/fm^3} for the Gaussian model. Borie \cite{bor2005a} finds \unit{-5.1975}{meV/fm^2} and \unit{0.0347}{meV/fm^3} for an exponential model, and \unit{0.0317}{meV/fm^3} for a Gaussian model, in reasonable agreement with the result presented here.

 In the same way, I evaluated the finite size correction to the fine structure.
 \begin{equation}
 \begin{split}
\label{eq:coulfsfsc}
\Delta E^{\mathrm{Coul.}} _{2p_{3/2}-2p_{1/2}}(R)&=
%checked July 22nd, 2012
8.41563570 \\
&- 0.00005192 R^2 \\
& +  1.1818650\times10^{-7} R^3 \\
& - 1.19528126\times10^{-9} R^4
 \, \mathrm{meV}.
\end{split}
\end{equation}
The constant term is in perfect agreement with Borie's value \unit{8.41564}{meV} \cite{bor2012} (Table 7).

It is interesting to explore at this stage the influence of the charge distribution shape on the Coulomb and vacuum polarization contribution.
Friar and Sick \cite{fas2005a} have evaluated the third Zemach moment from Eq. \eqref{eq:zem-rad3}, using the proton-electron scattering data available in 2005. Using a model-independent analysis, they find
\unit{ \left<r^3\right>_{(2)}=2.71\pm0.13}{fm^3}, leading to an energy shift of \unit{-0.0247\pm0.0012}{meV}.  
%Voir d'où ça vient: leading to shifts of  \unit{-0.0232}{meV}  \unit{-0.0212}{meV} respectively. 
Using the Fourier transform of the exponential \eqref{eq:exp-formfac} distribution in Eq. \eqref{eq:zem-rad3q}, I find 
\begin{equation}
\label{eq:dip-r3-2}
\left<r^3\right>_{(2)}=\frac{35 \sqrt{3} R^3}{16}\approx 3.789R^3,
\end{equation}
 showing that $\left<r^3\right>_{(2)}$ is proportional to $R^{3}$ and justifying the fit in $R^2$ and $R^3$ performed to derive the coefficients above.   Equation \eqref{eq:dip-r3-2} is in exact agreement with the result that can be obtained from Eq. (15) in \cite{ruj2010}, but Eq. (16) in the same work is not correct (the denominator should be 256, not 64). The value obtained by Friar and Sick corresponds  to $R=0.894$~fm. In that case our energy shift is \unit{-0.0250}{meV} in good agreement with the energy shift in Ref. \cite{fas2005a}.
 In the Gaussian model, I find
\begin{equation}
\left<r^3\right>_{(2)}=\frac{32 R^3}{3 \sqrt{3 \pi }}\approx 1.960R^3,
\end{equation}
leading to \unit{R=0.920}{fm} and an energy shift of \unit{-0.0256}{meV}, still in agreement. One can perform a more advanced calculation, using the experimental charge distribution from Ref. \cite{amt2007},  as given in \eqref{eq:form-fac-exper}. I find \unit{\left<r^3\right>=2.45}{fm^3}, significantly lower than Friar and Sick's value. This lead to \unit{R=0.864}{fm} for the exponential model and $R=0.889$~fm for the Gaussian model, providing  shifts of \unit{-0.0226}{meV} and \unit{-0.0232}{meV} respectively, in closer agreement to Borie's work. In a recent paper, De Rùjula \cite{ruj2010} claims that the discrepancy found between the charge radii obtained from hydrogen and muonic hydrogen could be due to the fact that theoretical calculations use too simple a dipole model to represent the nucleus. He builds a ``toy model'' composed of the sum of two dipole function corresponding to two resonances with different masses. In his model the third moment of the charge distribution is much higher than what is derived from a dipole model, enabling to mostly resolve the discrepancy between charge radii obtained from muonic and normal hydrogen.
He gets
\begin{equation}
\left<r^3\right>_{(2)}=36.6\pm7.3\approx 43 R^3,
\end{equation}
using $R$ from muonic hydrogen. I use the fit to the experimental form factor from Ref. \cite{amt2007} as given in Eq.\eqref{eq:form-fac-exper} to check the result from Ref. \cite{ruj2010} against  an experimental determination.
I obtain
\begin{equation}
\left<r^3\right>_{(2)}\approx 2.4485\approx 3.98 \times 0.850^3,
\end{equation}
very close to the dipole model value of Eq. \eqref{eq:dip-r3-2}. Using the recent MAMI experiment, combined with  data from \cite{amt2007}, Distler et al. \cite{dbw2011} obtain 
\begin{equation}
\left<r^3\right>_{(2)}\approx 2.85(8)\approx 4.18 \times 0.880^3,
\end{equation}
The conclusions from Ref. \cite{ruj2010}, which depend on an overly large third moment of the charge distribution are thus not supported by experiment.

%\begin{table}[tbp]
%\caption{Comparison of the coefficients for the finite nuclear size effect on the Vacuum polarization and Coulomb $2p_{1/2}$--$2s$ energy splitting }
%\begin{center}
%\begin{tabular}{ldd}
%\hline					
%\hline					
%Model	&	\multicolumn{1}{c}{$a$ (meV/fm$^2$)}	&	\multicolumn{1}{c}{$b$	(meV/fm$^3$)}\\
%\hline					
%Uniform	&	-5.2284	&	0.0313	\\
%Dipole	&	-5.2271	&	0.0353	\\
%Fermi	&	-5.2271	&	0.0324	\\
%Gauss	&	-5.2265	&	0.0328	\\
%Ref.\cite{bor2005a}, Dip.	&	-5.2248	&	0.0347	\\
%Ref.\cite{pac1999}	&	-5.225	&	0.0347	\\
%Ref.\cite{bor2005a}, Gauss	&	-5.2248	&	0.0317	\\\hline					
%\hline					
%\end{tabular}
%\end{center}
%\label{tab:vpfn}a
%\end{table}%

%\begin{figure}[htbp]
%\begin{center}
%\includegraphics[width=\columnwidth]{nuclear-size-dep-V11Dir}
%\caption{Dependence of  $\frac{\Delta E_{\textrm{V11FN}}}{R^2}$  as a function of $R$ in meV/fm$^2$ for different charge distribution models.}
%\label{fig:vpfn}
%\end{center}
%\end{figure}

\subsection{Finite size correction to the Uehling contribution}
For the vacuum polarization we obtain, for a point nucleus,
\begin{equation}
\Delta E^{11,\mathrm{pn}}_{2s_{1/2}-2p_{1/2}}= 205.028201\, \mathrm{meV},
\label{eq:vppnnum}
\end{equation}
 to be compared with \unit{205.0282}{meV} in Ref. \cite{bor2005a}. Pachucki \cite{pac1999} obtained \unit{205.0243}{meV} as the sum of the non-relativistic \unit{205.0074}{meV} and first order relativistic \unit{0.0169}{meV} corrections. If I calculate the difference between \eqref{eq:vppnnum} and Pachuki non-relativistic value, I obtain a difference of  \unit{0.0208076}{meV}. This is in excellent agreement with the value provided
 in Ref. \cite{kik2012}, Eq. (6), \unit{0.020843}{meV}. 
 
 To achieve this result we used the mesh parameters described in the previous section, and checked by varying them so that the results were stable within the decimal places provided here. For finite nuclei, I use the same parameters as in the previous section. Again changes in $r_0$ and $h$ do not change the final value. We get
\begin{equation}
\begin{split}
\Delta E^{11,\mathrm{fs}}_{2s_{1/2}-2p_{1/2}}(R)&=
% 205.0282071 - 0.02810417308 R^2  \\ 
%& + 0.0007016474313 \, R^3   \\
% &-0.00003007696961\, R^4   \\ 
% &-1.110588476\times10^{-6}R^5 \, \mathrm{meV}. \\
% updated May 20th 2012, New MMA file
%checked Juy 21st
205.0282076 - 0.02810970909 R^2 \\
&+ 0.0007111893365 R^3 \\
& -  0.00003572368803 R^4
 \label{eq:vp11fs}
\end{split}
\end{equation}
The constant term is in excellent agreement with the one in Eq. \eqref{eq:vppnnum}.
This result must be combined to Eq. \eqref{eq:coulfsr6log2} to obtain values that can be compared with the literature. I obtain:
\begin{equation}
\begin{split}
\Delta E^{11C,\mathrm{fs}}_{2s_{1/2}-2p_{1/2}}(R)&= 
%205.028219423 -5.22824684601 R^2 \\ 
%&+0.0369154093367 R^3 \\
%& -0.00112354084136 R^4  \\
%&+0.000430404343630 R^5 \\
%&-0.0000653864719937 R^6\, \mathrm{meV}.
% updated May 20th 2012, New MMA file
%205.0282076 - 5.227446248 R^2 \\
%&+ 0.03529257801 R^3 +  0.00007356191826 R^4 \\
%&+ 0.0002788380236 R^2 \log(R) \\
%&- 0.00004957597920 R^4 \log(R) \, \mathrm{meV}.
% July 21st, 2012
205.0282076 - 5.227446248 R^2 \\
&+ 0.03529257801 R^3 \\
& + 0.00007356191826 R^4 \\
& + 0.0002788380236 R^2 \log(R) \\
& -  0.00004957597920 R^4 \log(R)
\label{eq:vp11cfs} 
\end{split}
\end{equation}
The $R^2$ coefficient can be compared to the one in Ref. \cite{kik2012} Table III, which has the value \unit{-5.2254}{meVfm^{-2}}, which contains additional recoil corrections.
%The results for the proton size from  muonic hydrogen \cite{pana2010} and the 2010 CODATA adjustment \cite{CODATA2010}  are presented in Table \ref{tab:vpfnc}. The differences between Borie's calculations and the present one are at the level of 6\% of
%what would be required to explain the present discrepancy between hydrogen spectroscopy and electron-proton scattering results on one side and muonic hydrogen on the other side.

For the Uelhing correction to the fine structure, I obtain in the same way:
\begin{equation}
\begin{split}
\Delta E^{11,\mathrm{fs}}_{2p_{1/2}-2p_{3/2}}(R)&= 
%  0.00501579 \\
%&- 1.1608057\times10^{-7} R^2 \\
%&+  1.706858\times 10^{-9} R^3 
% july 22nd, 2012
0.0050157881 \\
& - 1.1662334\times10^{-7} R^2 \\
& + 2.2741334\times10^{-9} R^3  \\
& -  1.5308196\times10^{-10} R^4
\, \mathrm{meV}, 
\label{eq:vp11fsfs} 
\end{split}
\end{equation}
where the constant term is again in perfect agreement with Borie's value \unit{0.0050}{meV} \cite{bor2012} (Table 7).
%\begin{table}[tbp]
%\caption{Comparison of the sum of vacuum polarization and coulomb contribution calculations  for the $2p_{1/2}$--$2s$ energy interval for different proton sizes in a dipole model. Results are shown for the muonic hydrogen value \cite{pana2010} and the CODATA 2010 value \cite{CODATA2010}.}
%\begin{center} 
%\begin{ruledtabular}
%\begin{tabular}{ldd}
%$R$ (fm)	&	0.84184	&	0.8775	\\
%\hline
%direct calculation	&	201.34461	&	201.02690	\\
%Fit 3rd order	&	201.34462	&	201.02692	\\
%Fit 6th order	&	201.34461	&	201.02691	\\
%Borie	&	201.36367	&	201.04751	\\
%Diff.	&	-0.01906	&	-0.02061	\\\end{tabular}
%\end{ruledtabular}
%\end{center}
%\label{tab:vpfnc}
%\end{table}%

%We have used our numerical results for values of $R$ from \unit{0.7}{fm} to \unit{2.2}{fm} by step of \unit{0.1}{fm} to obtain $a$ and $b$. %Their values are given in Table \ref{tab:vpfn} and plotted on Fig.~\ref{fig:vpfn}. 

\subsection{Finite size correction to the Källén and Sabry contribution}
We can then evaluate the  Källén and Sabry  contribution $\Delta E^{21}_{2p_{1/2}}-\Delta E^{21}_{2s_{1/2}}$ using $V_{21}$ calculated following Sec. \ref{subsec:v21hp}, with good accuracy, using our numerical wavefunctions. For a point nucleus I obtain
\begin{equation}
 \Delta E^{21,\mathrm{pn}}_{2s_{1/2}-2p_{1/2}} = 1.508097\, \mathrm{meV},
\label{eq:kspn}
\end{equation}
in agreement with the result of Ref. \cite{pac1999}, \unit{1.5079}{meV}, and in excellent agreement with the one from Ref. \cite{bor2005a}, \unit{1.5081}{meV}. 
Using the wavefunctions calculated with the proton size, I can also evaluate the finite size correction to the Källén and Sabry contribution. A direct fit to the numerical data gives a result of the form 
\begin{equation}
%checked July 21st, 2012
\begin{split}
 \Delta E^{21,\mathrm{fs}}_{2s_{1/2}-2p_{1/2}}(R)&=1.508097 - 0.00021341293 R^2 \\
 &+ 7.3404895\times10^{-6} R^3  \\
 &- 5.0291143\times10^{-7} R^4 \, \mathrm{meV}.
\label{eq:ks21fs}
  \end{split}
\end{equation}
and 
\begin{equation}
\begin{split}
\label{eq:ks21fsfs}
 \Delta E^{21,\mathrm{fs}}_{2p_{1/2}-2p_{3/2}}(R)&= 
% 0.0000414300 \\
% & - 9.19105\times10^{-10} R^2 \\
% &+ 1.66897\times10^{-11} R^3
%updated July 22nd, 2012
0.0000414300 - 9.25489\times10^{-10} R^2 \\
&+ 2.33622\times10^{-11} R^3 \\
& -  1.80063\times10^{-12} R^4
 \, \mathrm{meV}.
  \end{split}
\end{equation}
for the fine structure, to be compared with \unit{0.00004}{meV} in Ref. \cite{bor2012}.

%\begin{table}[tbp]
%\caption{Finite size correction to the Källén and Sabry  contribution. Coefficients corresponds to Eq. \eqref{eq:ksfs}.}
%\begin{center}
%\begin{tabular}{lrrrr}
%\hline
%\hline
%Model	&	Uniform	&	Exponential	&	Fermi	&	Gaussian	\\
%\hline									
%$a$	&	-0.0002145	&	-0.0002145	&	-0.0002146	&	-0.0002145	\\
%$b$ &	0.0000078	&	0.0000086	&	0.0000082	&	0.0000083	\\
%$c$ 	&	-0.0000008	&	-0.0000009	&	-0.0000008	&	-0.0000009	\\
%\hline
%\hline
%\end{tabular}
%\end{center}
%\label{tab:v21fs}
%\end{table}%

\section{Higher-order vacuum polarization corrections}
\label{sec:hhvp}
\subsection{Higher-order vacuum polarization}
\label{subsec:hhv11}
\begin{figure}[tb]
	\centering
%====================
\includegraphics[width=0.7\columnwidth]{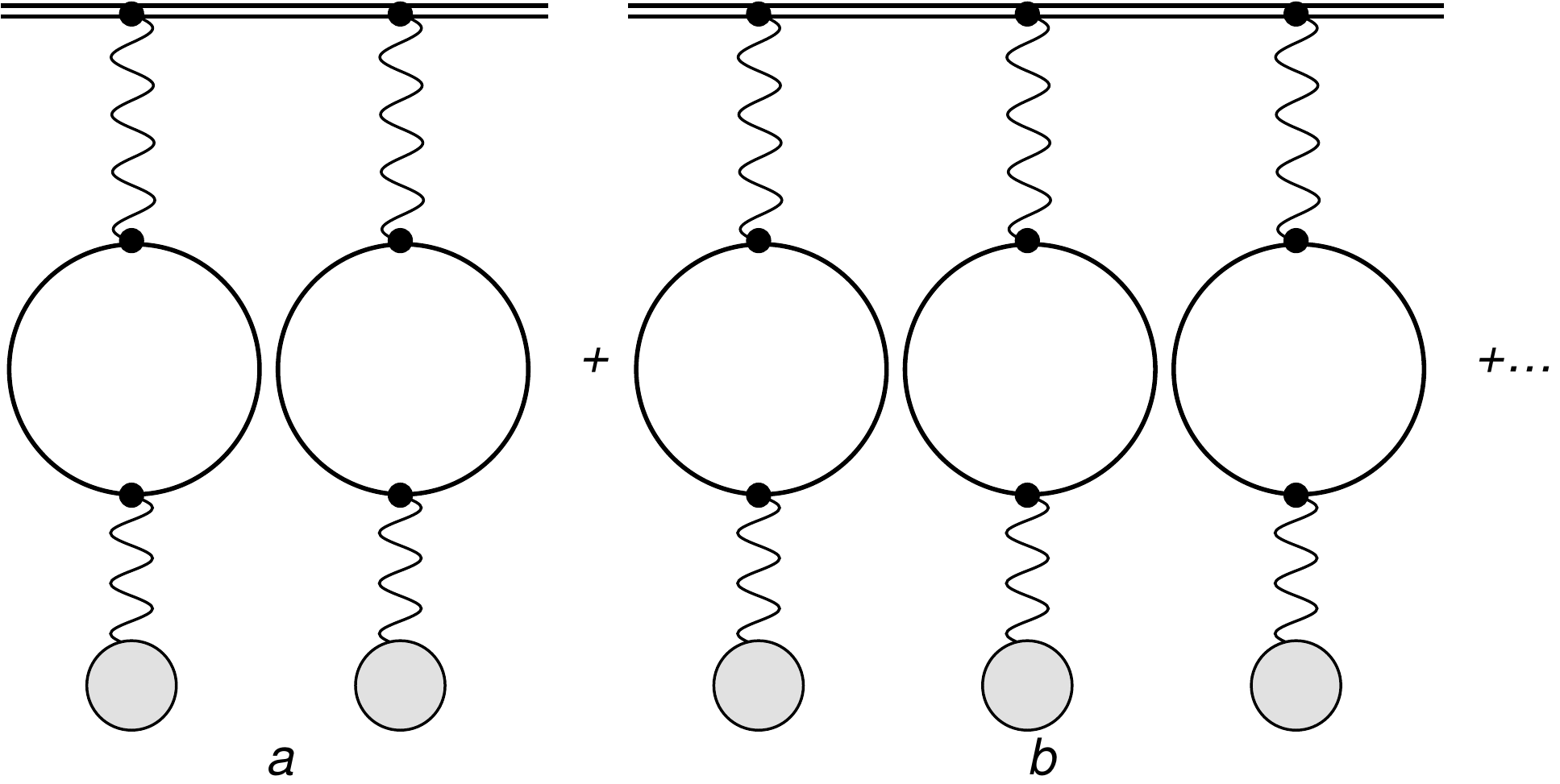}
	\caption[]{Feynman diagrams obtained when the Uehling potential is 
	added to the nuclear potential in the Dirac equation.  A double line represents a 
	bound electron wavefunction or propagator and a wavy line a 
	retarded photon propagator. The grey circles correspond to the interaction with the nucleus.
	Diagram (b) correspond to Fig. 5 and third term in Eq. (25) in Ref. \cite{kan1999}} \label{fig:vpsc1}
\end{figure}

The term named ``VP iteration'', which correspond to  Fig. \ref{fig:vpsc1}, is given by Eq. (215) of Ref.~\cite{egs2001}
\begin{equation}
  \label{eq:two-loop-vp}
  \Delta E_{\mathrm{VPVP}}(2s) = 0.01244 \frac{4}{9}\left(\frac{\alpha}{\pi}\right)^2 \left(Z\alpha\right)^2 \mu_{\mathrm{r}} c^2
\end{equation}
where $\mu_{\mathrm{r}}=$\unit{94.96446}{MeV} for muonic hydrogen (using \cite{mtn2008}).
This adds \unit{0.15086}{meV} to the Lamb-shift for muonic hydrogen. The Uehling potential under the form used in Sec.\ref{sec:vacpol} can be introduced in the potential of the Dirac equation \eqref{eq:dirac-diff} when solving it numerically. This amounts to get the exact solution with any number of vacuum polarization insertions as shown in Fig. \ref{fig:vpsc1}. The numerical methods that we used are described  in Ref. \cite{kla1977,bai2000}. Because of the Logarithmic dependence of the point nucleus Uehling potential at the origin, we do not calculate the iterated vacuum polarization directly for point nucleus. We instead calculate for different  mean square radii and charge distribution models, and fit the curves with $f(R)=a+b R^2+c R^3+dR^4$. All 4 models provides very similar values. The final value is 
\begin{equation}
\begin{split}
\label{eq:itervp}
\Delta E^{11,\mathrm{loop, fs}}_{2s_{1/2}-2p_{1/2}}(R)%&=0.1510170(3) -0.0000759(1) R^2\\
%&=0.1510212  \\
%& - 0.000095857 R^2\\
%& + 5.23555 \times 10^{-6} R^3
%updated July 21st, 2012
&=0.15102161\\
& - 0.000098409804 R^2 \\
&+ 7.9038238\times10^{-6} R^3\\
& -  7.2004764\times10^{-7} R^4
 \, \mathrm{meV}.
\end{split}
\end{equation}
The value calculated in Ref. \cite{pac1999} is \unit{0.1509}{meV} and the one in Ref.  \cite{bor2005a} is \unit{0.1510}{meV} in very good agreement with the present work. The method employed here provides in addition the proton size dependence for this correction, which was not calculated before. For the fine structure, I obtain in the same way
\begin{equation}
\begin{split}
\label{eq:itervpfs}
\Delta E^{11,\mathrm{loop, fs}}_{2p_{1/2}-2p_{3/2}}(R)&= 
%2.332\times10^{-6} \\
%&- 5.505\times10^{-10} R^2
% updated July, 22nd 2012
2.33197 \times 10^{-6}\\
& - 1.77101 \times 10^{-9} R^2 \\
&+ 9.36429 \times 10^{-10} R^3 \\
& -  1.79951 \times 10^{-10} R^4
 \, \mathrm{meV}.
\end{split}
\end{equation}

\subsection{Other higher-order Uehling correction}
\label{subsec:ho-diag}
\begin{figure}[tb]
	\centering
%====================
\includegraphics[height=50mm]{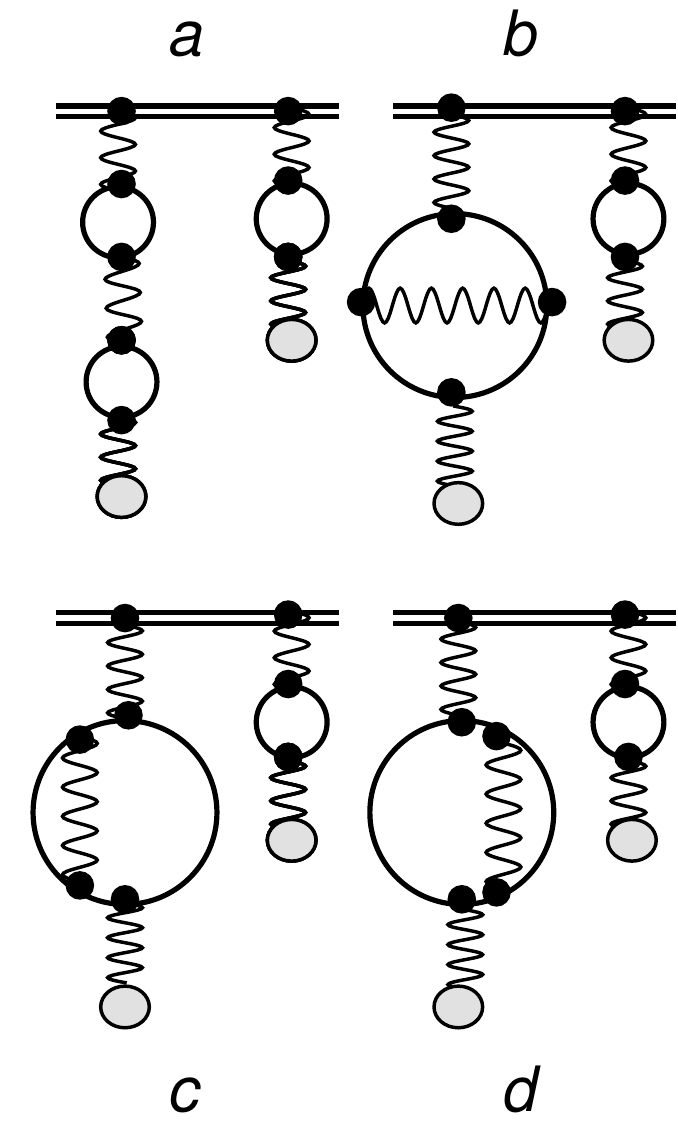}
	\caption[]{Lower order Feynman diagrams included in the Källén and Sabry $V_{21}(r)$ potential, when the Uehling potential is included in the differential equation. See Figs. \ref{fig:vp11} and \ref{fig:vpsc1} for explanation of symbols.
	Diagrams (a) and (b) exactly correspond to diagrams (a) and (b) in Fig. 5 and Eq. (25) in Ref. \cite{kan1999}} \label{fig:kas-v11}
\end{figure}

Since we include the vacuum polarization in the Dirac equation potential, all energies calculated by perturbation using the numerical wavefunction contains the 
contribution of higher-order diagrams where the external legs, which represent the wavefunction, can be replaced by a wavefunction and a bound propagator 
with one, or several vacuum polarization insertion. For example the  Källèn and Sabry correction calculated in this way, contains correction of the type 
presented in Fig. \ref{fig:kas-v11}.  This correction is given by 
%\begin{equation}
%\Delta E_{\mathrm{VP}\times \mathrm{KS}}=0.0021551 -0.0000012 R^2 \,\mathrm{meV}
%\label{eq:vp11xks}
%\end{equation}
 \begin{equation}
\begin{split}
%OK JUly 21st 2012
 \Delta E^{21\times 11,\mathrm{fs}}_{2s_{1/2}-2p_{1/2}}(R)&=0.0021552 \\
 &- 1.32976\times10^{-6} R^2 \\
 & + 9.4577\times10^{-8} R^3 \\
 & - 8.5185\times10^{-9} R^4 \,\textrm{meV},
\end{split}
\label{eq:vp11xks}
\end{equation}
with a $10^{-7}$ meV accuracy. This correction is part of the three-loop corrections form Ref. \cite{kan1999,ikk2009}. The diagrams in  Fig. \ref{fig:kas-v11} correspond to diagrams (a) and (b) of Fig.~5 in Ref. \cite{kan1999} and (e) (upper left) and (f) in Fig. 2 of Ref. \cite{kiks2010}. The sum of contributions of the diagram (a) and (b) is 0.00223, in good agreement with our all-order fully relativistic result. The three loop diagram Fig. 5 (c) Ref. \cite{kan1999} and Fig. 2 (g)  Ref. \cite{kiks2010} is included in the all-order contribution obtained by solving numerically the Dirac equation with the Uëlhing potential.
For the fine structure this correction is very small:
 \begin{equation}
\begin{split}
 \Delta E^{21\times 11,\mathrm{fs}}_{2p_{1/2}-2p_{3/2}}(R)&= 
 % 3.7575\times10^{-8} \\
% & - 3.3960\times10^{-12} R^2 \\
% &+ 1.5663\times10^{-13} R^3
% updated July 21st, 2012
3.75754\times 10^{-8}  \\
&- 3.49318\times 10^{-12} R^2 \\
& + 2.58244\times 10^{-13} R^3 \\
& - 2.74201\times 10^{-14} R^4
\,\textrm{meV}.
\end{split}
\label{eq:vp11xksfsfs}
\end{equation}

\subsection{Wichmann and Kroll correction}
\label{subsec:wak}
\begin{figure}[tb]
	\centering
%====================
\includegraphics[height=40mm]{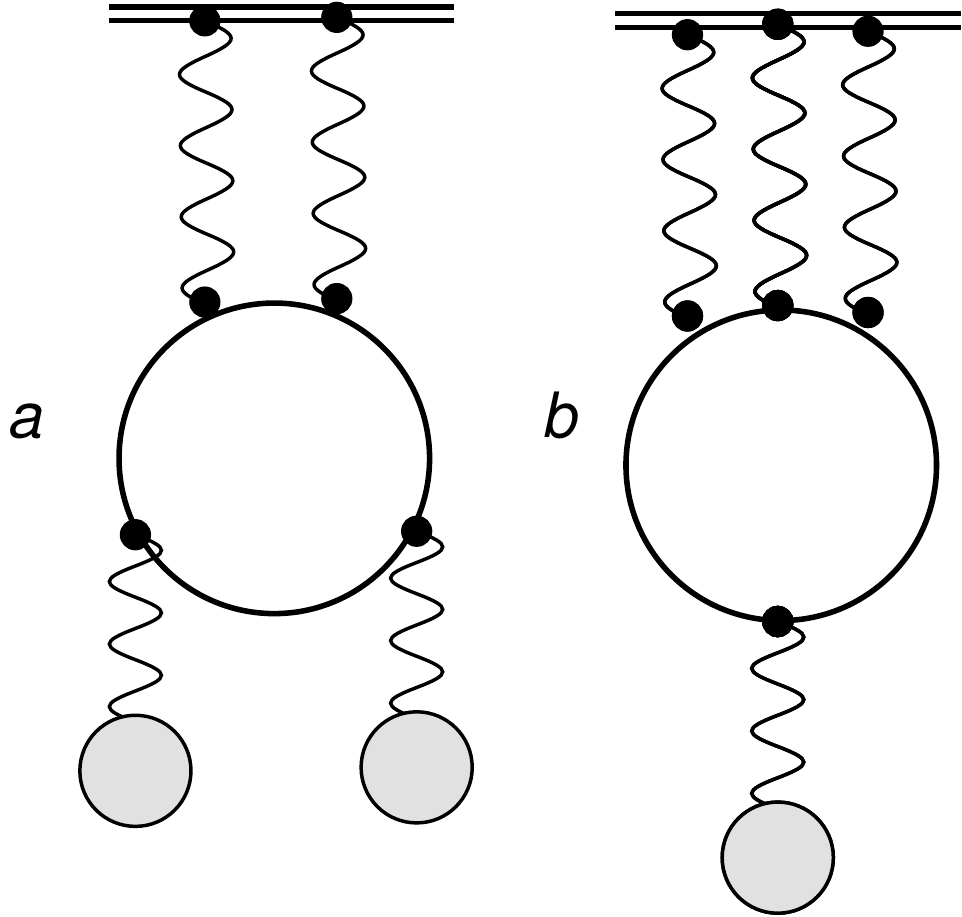}
	\caption[]{Feynman diagrams for the light-by-light scattering. See Figs. \ref{fig:vp11} and \ref{fig:vpsc1} for explanation of symbols.} \label{fig:lbl}
\end{figure}

We use the approximate potentials as presented in Refs. \cite{blo1972,hua1976} to evaluate the Wichmann and Kroll \cite{wak1956} $V_{13}$ correction to the Uehling potential. The corresponding diagram is shown on Fig. \ref{fig:vp11} (b). This contribution is given together with the light-by-light scattering diagrams of Fig. \ref{fig:lbl} in Refs. \cite{kan1999,kiks2010,kkis2010}. We find for a point nucleus, the exact value, and a size correction, given by
\begin{equation}
\begin{split}
\Delta E^{13,\mathrm{fs}}_{2s_{1/2}-2p_{1/2}}(R)&=
%-0.001017063 \\
%&+ 5.54828 \times 10^{-8} R^2 \\
%& - 5.85327\times10^{-10} R^3
%updated July 22nd 2012
-0.0010170628 \\
&+ 5.5414179\times 10^{-8} R^2\\
& - 5.1356872 \times 10^{-10} R^3 \\
&  - 1.9364450 \times10^{-11} R^4 
  \,\mathrm{meV},
\label{eq:wk} 
\end{split}
\end{equation}
to be compared to the value given in Ref. \cite{kiks2010} (Table III)  of \unit{-0.001018(4)}{meV}. In the lowest order approximation, the diagram on  Fig. \ref{fig:lbl}(a) provides an energy shift of $\Delta E_{13}/Z^2$  \cite{kiks2010,kkis2010}. 
For the fine structure, this correction is comparable to the contribution from Eq. \eqref{eq:vp11xksfsfs}:
\begin{equation}
\begin{split}
\Delta E^{13,\mathrm{fs}}_{2p_{1/2}-2p_{3/2}}(R)&= 
-4.21088\times 10^{-8} \\
 &+ 4.41081 \times 10^{-13} R^2 \\
 &- 8.49036 \times 10^{-15} R^3 \\
 &+  1.81474 \times 10^{-15} R^4
  \,\mathrm{meV}.
\label{eq:wkfs} 
\end{split}
\end{equation}

%\begin{figure}[tb]
%	\centering
%%====================
%\includegraphics[height=40 mm]{wich-kroll}
%	\caption[]{Feynman diagrams for the lower-order Wichmann and Kroll $V_{13}(r)$ potential. See Fig. \ref{fig:vp11} for explanation of symbols.} \label{fig:v13}
%\end{figure}

\subsection{Muon radiative corrections}
\label{subsec:muon-radiat}
\subsubsection{Muon self-energy}
\label{subsubsec:muonse}
\begin{figure}[tb]
	\centering
%====================
\includegraphics[width=0.5\columnwidth]{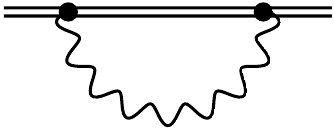}
	\caption[]{Feynman diagrams for the muon self-energy. See Figs. \ref{fig:vp11} and \ref{fig:vpsc1} for explanation of symbols.} \label{fig:se}
\end{figure}

Highly accurate self-energy values for electronic atoms and point nucleus are known from Ref.\cite{jms1999}. The self-energy correction, represented in Fig. \ref{fig:se} is conveniently expressed by the  slowly varying  function $F(Z\alpha)$ defined by
\begin{equation}
\label{eq:scaled-SE} 
\Delta E_{SE}= \frac{\alpha}{\pi}\frac{ (Z\alpha)^4}{n^3} mc^2 F(Z\alpha),
\end{equation}
where $m$ is the particle mass.

The recoil corrections to $ F(Z\alpha)$ are described in detail 
in \cite{mtn2008}. The dependence in the reduced mass has to be included leading to the following expressions, specialized for the $n=2$ shells:
\begin{eqnarray}
\label{muon-se-2S}
\Delta E_{\mu SE,2S}&=&\frac{\alpha}{\pi}\frac{(Z\alpha)^4}{8}\left(\frac{\mu_{\mathrm{r}}}{m_{\rmssmu}}\right)^3 m_{\rmssmu}c^2\left(-\frac{4}{3}\ln k_0(2S)+\frac{10}{9}\right. \nonumber \\
&&+\frac{4}{3}\ln\left(\frac{m_{\rmssmu}}{\alpha^2\mu_{\mathrm{r}}}\right)+\left(\frac{139}{32}-2 \ln 2\right)\pi\alpha \nonumber \\
&& + \left(\frac{67}{30}+\frac{16\ln 2}{3}\right)\ln\left(\frac{m_{\rmssmu}}{\alpha^2\mu_{\mathrm{r}}}\right)\alpha^2 \nonumber \\
&&-\left(\ln\left(\frac{m_{\rmssmu}}{\alpha^2\mu_{\mathrm{r}}}\right)\right)^2\alpha^2\nonumber \\
&&\left. +\alpha^2G_{2s}(\alpha)\right),
\end{eqnarray}
\begin{eqnarray}
\label{muon-se-2p1}
\Delta E_{\mu SE,2p_{1/2}}&=&\frac{\alpha}{\pi}\frac{(Z\alpha)^4}{8}\left(\frac{\mu_{\mathrm{r}}}{m_{\rmssmu}}\right)^3m_{\rmssmu}\left(-\frac{4}{3}\ln k_0(2P)-\frac{1}{6}\left(\frac{m_{\rmssmu}}{\mu_{\mathrm{r}}}\right)\right. \nonumber \\
&&\left.+\frac{103}{180}\ln\left(\frac{m_{\rmssmu}}{\alpha^2\mu_{\mathrm{r}}}\right)\alpha^2+\alpha^2G_{2p_{1/2}}(\alpha)\right),
\end{eqnarray}
and
\begin{eqnarray}
\label{muon-se-2p3}
\Delta E_{\mu SE,2p_{3/2}}&=&\frac{\alpha}{\pi}\frac{(Z\alpha)^4}{8}\left(\frac{\mu_{\mathrm{r}}}{m_{\rmssmu}}\right)^3m_{\rmssmu}\left(-\frac{4}{3}\ln k_0(2P)+\frac{1}{12}\left(\frac{m_{\rmssmu}}{\mu_{\mathrm{r}}}\right)\right. \nonumber \\
&&\left.+\frac{29}{90}\ln\left(\frac{m_{\rmssmu}}{\alpha^2\mu_{\mathrm{r}}}\right)\alpha^2+\alpha^2G_{2p_{3/2}}(\alpha)\right).
\end{eqnarray}
The Bethe logarithms are given by $\ln k_0(2S)=2.811 769 893$ and $\ln k_0(2P)=-0.030 016 709$ \cite{das1990}. The 
the remainders are given by $G_{2s}(\alpha)=-31.185 150(90)$, $G_{2p_{1/2}}(\alpha)=-0.973 50(20)$ and $G_{2p_{3/2}}(\alpha)=-0.486 50(20)$ \cite{jms1999,jms2001}. One then gets the exact 
muon self-energy for each state. For the $2s$ state, this gives \unit{0.675150}{meV} instead of \unit{0.675389}{meV}. For the $2p_{1/2}$ I get \unit{-0.00916882}{meV} and for $2p_{3/2}$, \unit{0.008393377}{meV} in place of
\unit{0.01424054}{meV} and \unit{-0.00332838}{meV} respectively, if one would use only the low order $A_{40}$ term.

The finite size correction is given by perturbation theory \cite{mtn2008} Eq. (54)
\begin{equation}
  \label{eq:se-fs}
  E_{SE-NS}(R,Z\alpha)=\left(4 \ln 2 -\frac{23}{4}\right)\alpha (Z\alpha){\cal E}_{NS}(R,Z\alpha),
\end{equation}
where (\cite{mtn2008} Eq. (51))
\begin{equation}
  \label{eq:coul-fs}
  {\cal E}_{NS}(R,Z\alpha) = \frac{2}{3}\left(\frac{\mu_{\mathrm{r}}}{m_{\rmssmu}}\right)^3\frac{(Z\alpha)^2}{n^3}m_{\rmssmu}\left(\frac{Z\alpha R}{\lambdabar_{\textrm{C}}}\right)^2,
\end{equation}
is the lowest-order finite nuclear size correction to the Coulomb energy. Here $\lambdabar_{\textrm{C}}=1.867594282\,\textrm{fm}$ is the muon Compton wavelength.
Equation \eqref{eq:coul-fs} provides ${\cal E}_{NS}(R)=5.19745R^2$ for muonic hydrogen in agreement with Refs. \cite{pac1999,bor2005a}.
%One can rewrite Eqs. \eqref{eq:se-fs} and \eqref{eq:coul-fs} in term of the reduced function as 
%\begin{eqnarray}
%  \label{eq:FS-lo}
%  F_{NS}^{lo}(R,Z\alpha) &=&\left(4 \ln 2 -\frac{23}{4}\right) \frac{2\pi Z\alpha}{3}\left(\frac{\mu_{\mathrm{r}}}{m_{\rmssmu}}\right)^3\left(\frac{ R}{\lambdabar_{\textrm{C}}}\right)^2, \nonumber \\
%  &=& -0.0094726 R^2
%\end{eqnarray}

The self-energy correction to the Lamb shift with finite-size correction is then
\begin{equation}
\label{eq:muon-LS-se-fs}
 \Delta E^{SE,\mathrm{fs}}_{2p_{1/2}}-\Delta E^{SE,\mathrm{fs}}_{2s_{1/2}}(R)=-0.68431882+0.000824 R^2\,\textrm{meV},
\end{equation}
and to the fine structure:
\begin{equation}
\label{eq:muon-FS-se-fs}
 \Delta E^{SE,\mathrm{fs}}_{2p_{3/2}}-\Delta E^{SE,\mathrm{fs}}_{2p_{1/2}}=0.017562197\,\textrm{meV}.
\end{equation}
It should be noted that in Ref. \cite{bor2005a}, the $R^2$-dependent part of the $2s$ self-energy is much larger than what is given in Eq. \eqref{eq:muon-LS-se-fs}. This value was checked independently by using an all-order calculation with finite size, following the work of Mohr and Soff \cite{mas1993}. The results of this calculation agree reasonably well with  Eq. \eqref{eq:muon-LS-se-fs} and is given by  \cite{iam2012}
\begin{eqnarray}
  \label{eq:se-fs-H}
  E_{SE-NS}^H(Z=1,R)&=&-0.68431882 \nonumber \\
  %-0.00107 R^2 + 0.00035R^3\nonumber \\             &&  -0.00007 R^4\, \textrm{meV}.
%  -0.001677053 R^2 \nonumber \\
%&&  + 0.0009996784R^3 \nonumber \\
%&&- 0.0004785372 R^4 \nonumber \\
%&&+ 0.00008785574 R^5
% Corrected 1/11/2012 for recoil 
&& + 0.0012176389 R^2  \nonumber \\
&&- 0.00072582511 R^3  \nonumber \\ 
&& + 0.00034744609  R^4  \nonumber \\
&& - 0.000063788419 R^5
\,\textrm{meV}.
\end{eqnarray}

\subsubsection{Muon loop vacuum polarization}
\label{subsubsec:muonvp}
The vacuum polarization due to the creation of virtual muon pairs is represented by the same diagram \ref{fig:vp11} (a) and same equations \eqref{eq:v11} as vacuum polarization due to electron-positron pairs, replacing the electron Compton wavelength by the muon one.
For $S$ states, it is given by
\cite{mtn2008} Eq. (27),\cite{pac1996} Eq. (32)
\begin{equation}
\label{eq:muon-loop-vp}
E_{\mu VP}(ns)=-\frac{\alpha (\alpha Z)^4}{\pi n^3}\left(-\frac{4}{15}+\pi \alpha\frac{5}{48}\right)\left(\frac{\mu_{\mathrm{r}}}{m_{\rmssmu}}\right)^3m_{\rmssmu}c^2,
\end{equation}
in which higher order terms in $Z\alpha$ have been neglected. For the $2s$ Lamb shift in muonic hydrogen it gives \unit{0.01669}{meV} and is included in Refs. \cite{pac1999,bor2005a} and \cite{jen2011} Eq. (2.29) for the first $\alpha$ correction.
 As it is a sizable contribution, and the muon Compton wavelength, which represent the scale of QED corrections for muons
is of the order of the finite nuclear size (\unit{1.9}{fm}), one could expect a non-negligible finite size contribution.
Using the numerical procedure described in Sec. \ref{sec:vacpol}, replacing the electron Compton wavelength by the muon one in Eq. \eqref{eq:v11},  I obtain
\begin{equation}
\begin{split}
%E_{\mu VP}(2s) &= 0.00000180R^4 - 0.00003625R^2 + 0.01670254 \,\textrm{meV},\\
 \Delta E^{\mu11,\mathrm{fs}}_{2s_{1/2}-2p_{1/2}}(R)&=
%  0.0167133 \\
%& - 0.000064995 R^2 \\
%& + 0.000025203 R^3\\
%&  - 4.0589\times10^{-6} R^4
% updated July 21st 2012
0.01671487464 \\
& - 0.00005279721702 R^2\\
&+ 0.00001269866912 R^3 \\
&-  5.360546098\times10^{-6} R^4 \\
&+ 0.00001717649157 R^2 \log(R)  \\
&+  2.047113814\times10^{-6} R^4 \log(R)
 \,\textrm{meV},
\end{split}
\label{eq:muonvpfs}
\end{equation}
where the constant term is in excellent agreement with \eqref{eq:muon-loop-vp} and the $R$ dependence explicit.
From Ref. \cite{mtn2008}, Eq. (55), one obtains
\begin{equation}
\label{eq:pert-vp-fn-size}
E_{\mu VP}^{\mathrm{fs}}(ns)=\frac{3}{4}\alpha (Z\alpha){\cal E}_{NS}(R,\alpha)=0.00020758 R^2\,\textrm{meV} 
\end{equation}
for the $2s$ level. This term is about 4 times larger than the numerical coefficient for $R^2$ in Eq. \eqref{eq:muonvpfs}. 

Using the  wavefunction evaluated with the Uehling potential in the Dirac equation, I also obtain the value of the sum of diagrams with one muon vacuum polarization loop and any number of electron loops on each side, as in Fig. \ref{fig:vpsc1}, with one loop being a muon loop:
\begin{equation}
\begin{split}
%E_{\mu VP}(2s) &= 0.00000180R^4 - 0.00003625R^2 + 0.01670254 \,\textrm{meV},\\
 \Delta E^{\mu11\times 11,\mathrm{fs}}_{2s_{1/2}-2p_{1/2}}(R)&=
%  0.00005346 \\
% &- 7.6035\times10^{-7} R^2 \\
% &+ 3.2548\times10^{-7} R^3\\
% & - 5.4392\times10^{-8} R^4
% updated to mach MMA notebook, with log terms
0.00005348857 \\
& - 5.358667885\times10^{-7} R^2 \\
& + 6.495888541\times10^{-8} R^3 \\
& - 4.287455607\times10^{-8} R^4  \\
&+ 2.822488184\times10^{-7} R^2 \log(R) \\
&+  1.739448734\times10^{-8} R^4 \log(R) \\
& \,\textrm{meV}.
\end{split}
\label{eq:muonvpfsfs}
\end{equation}
This muonic vacuum polarization  is a  small contribution to the fine structure
\begin{equation}
\begin{split}
%E_{\mu VP}(2s) &= 0.00000180R^4 - 0.00003625R^2 + 0.01670254 \,\textrm{meV},\\
 \Delta E^{\mu11\times 11,\mathrm{fs}}_{2p_{1/2}-2p_{3/2}}(R)&= 
 % June 12, 2012
 1.67794\times10^{-7} \\
 &- 2.10861\times10^{-10} R^2 \\
&  + 8.51427\times10^{-11} R^3 \\
&   - 1.38884\times10^{-11} R^4
\,\textrm{meV}.
\end{split}
\label{eq:muonvpfsfs-fs}
\end{equation}

%%%================= RECOIL ====================
\section{Evaluation of the recoil corrections}
\label{sec:recoil}
\begin{figure}[tb]
	\centering
%====================
\includegraphics[width=0.7\columnwidth]{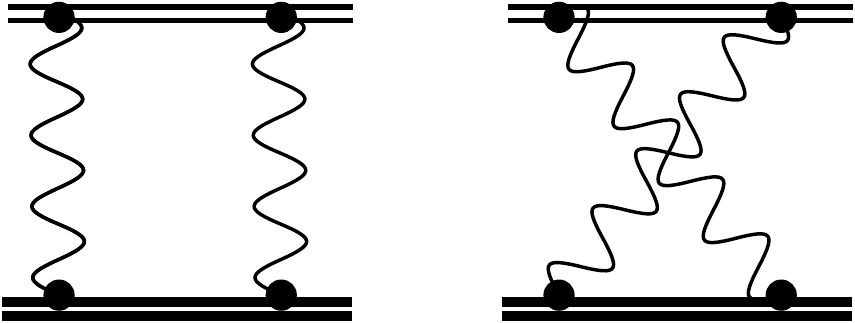}
	\caption[]{Feynman diagrams corresponding to  the Relativistic recoil correction \eqref{eq:rr1}. The heavy double line represents the proton wave function or propagator. The other symbols are explained in Fig. \ref{fig:vp11}. } \label{fig:rr1}
\end{figure}

The relativistic treatment of recoil corrections is described in, e.g., \cite{mtn2008}, Eq. (10).
The analytic solution of the Dirac equation for a point nucleus and a particle of mass $m$ is given by
\begin{equation}
E_{\mathrm{D}}= m c^2 f(n,j)
\label{eq:dirac-point}
\end{equation}
with 
\begin{equation}
f(n,j)= \frac{1}{\sqrt{1+\frac{(Z\alpha)^2}{\left(n-j+\frac{1}{2}+\sqrt{\left(j+\frac{1}{2}\right)^2-(Z\alpha)^2}\right)^2}}}.
\end{equation}
The recoil can then be included by evaluating \cite{bag1955,say1990}
\begin{eqnarray}
 E_M &=& Mc^2+\mu_{\mathrm{r}}c^2\left(f(n,j)-1\right) \nonumber \\
 && +\frac{\left[f(n,j)-1\right]^2\mu_{\mathrm{r}}^2c^2}{2M}  \label{eq:rec1} \\
 && +\frac{1-\delta_{l,0}}{\kappa(2l+1)}\frac{(Z\alpha)^4\mu_{\mathrm{r}}^3c^2}{2n^3M\rsp^2}  \label{eq:rec2},
\end{eqnarray}
where $M=m_{\rmssmu}+M\rsp$. If one expands the previous equation in power of $(Z\alpha)$, one would find that the terms of order up to $(Z\alpha)^4$ are identical to what is given in Ref. \cite{bag1955}. We also compared the numerical results from our numerical approach for point nucleus, as described in Sec. \ref{subsec:dirac} to what can be obtained by using directly \eqref{eq:dirac-point} and find excellent agreement. Below, we will make exclusive use of the direct numerical evaluation of the Dirac equation.
\
The relativistic corrections to Eq. \eqref{eq:rec2} associated with motion of the nucleus are called relativistic-recoil correction. The correction to order $(Z\alpha)^5$ and to all orders in $m_{\rmssmu}/M\rsp$ is given by \cite{eri1977,say1990,egs2007,mtn2008}
\begin{eqnarray}
E_{\mathrm{RR}}^5(n,l)&=&\frac{\mu_{\mathrm{r}}^3c^2}{m_{\rmssmu}M\rsp}\frac{(Z\alpha)^5}{\pi n^3}
\Bigg\{
\frac{\delta_{l,0}}{3}\ln\frac{1}{(Z\alpha)^2}-\frac{8}{3}\ln k_0(l,n) \nonumber \\
&& -\frac{\delta_{l,0}}{9} -\frac{7}{3}a_{n,l}-\frac{2\delta_{l,0}}{M\rsp^2-m_{\rmssmu}^2}  \nonumber \\
&& \times \left[M\rsp^2\ln\left(\frac{m_{\rmssmu}}{\mu_{\mathrm{r}}}\right)-m_{\rmssmu}^2\ln\left(\frac{M\rsp}{\mu_{\mathrm{r}}}\right)
\right]
\Bigg\}
\label{eq:rr1}
\end{eqnarray}
where 
\begin{eqnarray}
a_{n,l}&=&-2\left[\ln\left(\frac{2}{n}+\sum_{i=1}^{n}\frac{1}{i}\right)+1-\frac{1}{2n}\right]\delta_{l,0}\nonumber \\
&&+\frac{1-\delta_{l,0}}{l(l+1)(2l+1)}.
\label{eq:rr1-aux}
\end{eqnarray}
This correction corresponds to the diagrams in Fig. \ref{fig:rr1}. 

The next order of the relativistic recoil corrections is given for $s$ states  by
\begin{eqnarray}
E_{\mathrm{RR}}^6(ns)&=&\frac{m_{\rmssmu}}{M\rsp}\frac{(Z\alpha)^6}{n^3}m_{\rmssmu}c^2
\Bigg\{4\ln 2 -\frac{7}{2} \nonumber \\
&&-\frac{11}{60\pi}\ln\frac{1}{(Z\alpha)^2}
\Bigg\},
\label{eq:rr2s}
\end{eqnarray}
and for $l\ge1$ states by
\begin{eqnarray}
E_{\mathrm{RR}}^6(nl)&=&\frac{m_{\rmssmu}}{M\rsp}\frac{(Z\alpha)^6}{n^3}m_{\rmssmu}c^2
\Bigg\{\left[3-\frac{l(l+1)}{n^2}\right] \nonumber \\
&&\times\frac{2}{\left(4l^2-1\right)(2l+3)}
\Bigg\}.
\label{eq:rr2nl}
\end{eqnarray}
Using Eqs. \eqref{eq:rec1} to \eqref{eq:rr2nl}, I obtain
\begin{equation}
\begin{split}
\label{eq:recoil-2s-2p1}
 \Delta E^{\mathrm{Rec.}}_{2s_{1/2}-2p_{1/2}}&=0+0.0574706-0.0449705 \\
  &=0.0125001\textrm{meV},
 \end{split}
\end{equation}
and to the fine structure:
\begin{equation}
\label{eq:recoil-2p1-2p3}
\begin{split}
 \Delta E^{\mathrm{Rec.}}_{2p_{1/2}-2p_{3/2}}&= 0.0000051-0.0862059+0\\
 &=-0.0862008\,\textrm{meV}.
 \end{split}
\end{equation}
This is in excellent agreement with the results from Ref. \cite{bor2012}.

%%%================== HFS =======================
\section{Evaluation of some all-order hyperfine structure corrections}
\label{sec:hfs}
The expression of the hyperfine magnetic dipole operator can be written as 
\begin{equation}
\label{eq:HFS-hamilt-m1}
H_{hfs}= -e c \boldsymbol{\alpha}\cdot \boldsymbol{A}(\boldsymbol{r})=-e c \boldsymbol{\alpha}\cdot \boldsymbol{A}(\boldsymbol{r}),
\end{equation}
 with
  \begin{equation}
\label{eq:mag-moment-field}
 \boldsymbol{A}(\boldsymbol{r}) = \frac{\mu_0}{4\pi} \frac{\boldsymbol{\mu} \times \boldsymbol{r}}{r^3},
\end{equation}
 where $\boldsymbol{\mu}$ is the nuclear magnetic moment and we have assumed a magnetic moment distribution of a point particle for the nucleus. It is convenient to express $H_{hfs}$ using vector spherical harmonics.
On obtains \cite{sch1955,lar1974,cac1985,joh2007}
\begin{equation}
H_{hfs}=\boldsymbol{M}^1\cdot \boldsymbol{T}^1, 
\end{equation}
where 
\begin{equation}
\label{eq:t1-hfs-def}
\boldsymbol{T}^1\left(\boldsymbol{r}\right)=-i e \sqrt{\frac{8 \pi}{3} }\frac{\boldsymbol{\alpha} \cdot \boldsymbol{Y}_{1q}^{(0)}\left(\hat{\boldsymbol{r}}\right)}{r^2},
\end{equation}
and $\boldsymbol{M}^1$ representing the magnetic moment operator from the nucleus. The operator $\boldsymbol{T}^1$ acts only on the bound particle coordinates.
The vector spherical harmonic $\boldsymbol{Y}_{1q}^{(0)}\left(\hat{\boldsymbol{r}}\right)$ is an eigenfunction of $\boldsymbol{J}^2$ and $J_z$, defined as \cite{jud1963,lar1974,beg1985,vmk1988,joh2007}
\begin{equation}
\begin{split}
\label{eq:vect-sphe-harm}
 \boldsymbol{Y}_{1q}^{(0)}\left(\hat{\boldsymbol{r}}\right)&=\boldsymbol{Y}_{11q}\left(\hat{\boldsymbol{r}}\right) = \sum_{\sigma} C\left(1,1,1;q-\sigma,\sigma,q\right)Y_{1,q-\sigma}\left(\hat{\boldsymbol{r}}\right)\boldsymbol{\xi}_{\sigma} 
 \end{split}
\end{equation}
where $ C\left(j_1,j_2,j;m_1,m_2,m\right)$ is a Clebsh-Gordan coefficient, $Y_{1,q}$  are scalar spherical harmonic and $\boldsymbol{\xi}_{\sigma}$ are eigenvectors of $s^2$ and $s_z$, the spin 1 matrices \cite{jud1963,lar1974,beg1985,vmk1988,joh2007}. The reduction to radial and angular integrals is presented in various works \cite{lar1974,cac1985,joh2007}.
In heavy atoms, the hyperfine structure correction due to the magnetic moment contribution is usually calculated for a finite charge distribution, but a point magnetic dipole moment (see, e.g., \cite{lar1974,cac1985}). When matrix elements non-diagonal in $J$ are needed, one can use \cite{ipm1989} for a one-particle atom
\begin{equation}
  \label{eq:hfs-mcdf}
  \Delta E^{HFS}_{\textrm{M1}}=A \frac{g\alpha}{2M_p}  \int_{0}^{\infty} dr \frac{P_1(r) Q_2(r)+P_2(r)Q_1(r)}{r^2}, 
\end{equation}
where $g=\mu_p/2=2.792847356$ for the proton, is the anomalous magnetic moment, $A$ is an angular coefficient 
\begin{equation}
\begin{split}
  \label{eq:ang}
  A& =(-1)^{I+j_1+F}\frac{\left\{
      \begin{array}{ccc}
        I & j_1 & F \\
        j_2& I  & k \\
      \end{array}
\right\}
}{
\left(
  \begin{array}{ccc}
    I & 1 & I \\
    -I & 0  &I \\
  \end{array}
\right) 
} \\
&\qquad \times(-1)^{J_1-\frac{1}{2}}\sqrt{\left(2J_1+1\right)\left(2J_2+1\right)}
% eliminated by Wigner-Eckart
%\left(
 % \begin{array}{ccc}
 %   j_1& k & j_2 \\
 %   -m_1 & q  &m_2 \\
 % \end{array}
%\right)  \\
%& \qquad \times
\left(
  \begin{array}{ccc}
   j_1& 1 & j_2 \\
   \frac{1}{2} & 0  &-\frac{1}{2} \\
    \end{array}
    \right)
 \pi\left(l_1,k,l_2\right) ,
\end{split}
\end{equation}
where $\pi\left(l_1,k,l_2\right)=0$ if $l_1+l_2+1$ is odd and 1 otherwise. The $j_i$ are the total angular momentum of the $i$ state for the bound particle, $l_i$ are orbital angular momentum, $I$ is 
the nuclear spin, $k$ the multipole order ($k=1$ for the magnetic dipole contribution described in Eq. \eqref{eq:hfs-mcdf}) and $F$ the total angular momentum of the atom. The difference between $\Delta E^{HFS}$ values calculated with a finite or point nuclear charge contribution is called the Breit-Rosenthal correction \cite{rab1932}.

 To consider a finite magnetic moment distribution, one uses the Bohr-Weisskopf correction \cite{baw1950}. The correction can be written \cite{fwdm1980}
\begin{equation}
  \label{eq:bw-mcdf}
\begin{split}
  \Delta E^{BW}& =- A \frac{g\alpha}{2M_p} \int_{0}^{\infty} dr_n  r_n^2 \mu (r_n) \\
  & \quad \times \int_{0}^{r_n} dr \frac{P_1(r) Q_2(r)+P_2(r)Q_1(r)}{r^2}, 
  \end{split}
\end{equation}
where the magnetic moment density $\mu (r_n)$ is normalized as 
\begin{equation}
  \label{eq:mu-norm}
  \int_{0}^{\infty } dr_n  r_n^2 \mu (r_n) =1.
\end{equation}

Borie and Rinker \cite{bar1982},  write the total diagonal hyperfine energy correction for a muonic atom as
\begin{eqnarray}
  \label{eq:hfs-bor-rink}
  \Delta E_{i,j}&=&  \frac{4\pi \kappa \left(F(F+1)-I(I+1)-j(j+1)\right)}{\kappa^2-\frac{1}{4}}\frac{g\alpha}{2M_p} \nonumber \\
 && \times \int_{0}^{\infty} dr \frac{P_1(r) Q_2(r)}{r^2}  \int_{0}^{r} dr_n r_n^2  \mu_{BR} (r_n).
\end{eqnarray}
where the normalization is different:
\begin{equation}
  \label{eq:mu-norm-BR}
  \int_{0}^{\infty } d^3r_n   \mu_{BR} (r_n) = 4\pi \int_{0}^{\infty } dr_n r_n^2  \mu_{BR}(r_n)= 1.
\end{equation}
This means that $\mu_{BR}(r)=\mu(r)/(4\pi)$. Evaluation of the Wigner 3J and 6J symbols in \eqref{eq:ang} give the same angular factor than in Eq. \eqref{eq:hfs-mcdf}.

The equivalence of the two formalism can be easily checked:
starting from \eqref{eq:hfs-bor-rink} and droping the angular factors, we get, doing an integration by part
\begin{multline}
  \label{eq:part}
   \int_{0}^{\infty} dr \frac{P_1(r) Q_2(r)}{r^2}  \int_{0}^{r} dr_n r_n^2  \mu (r_n) \\
=\left[  \int_{0}^{r} dr_n r_n^2  \mu (r_n)\int_{0}^{r} dt\frac{P_1(t) Q_2(t)}{t^2}\right]_{0}^{\infty} \\
-\int_{0}^{\infty} dr_n  r_n^2 \mu (r_n) \int_{0}^{r_n} dr \frac{P_1(r) Q_2(r)}{r^2} \\
= \int_{0}^{\infty} dr \frac{P_1(r) Q_2(r)}{r^2}\\
-\int_{0}^{\infty} dr_n  r_n^2 \mu (r_n) \int_{0}^{r_n} dr \frac{P_1(r) Q_2(r)}{r^2},
 \end{multline}
where we have used \eqref{eq:mu-norm}. We thus find that the formula in Borie and Rinker represents the \emph{full} hyperfine structure
correction, including the Bohr-Weisskopf part.

In 1956, Zemach  \cite{zem1956} calculated the fine structure energy of hydrogen, including recoil effects. He showed that in first order in the finite size, the HFS depends on the charge \emph{and} magnetic distribution moments only through the Zemacs's form factor defined in Eq. \eqref{eq:em-form-factor}. The proton is assumed to be at the origin of coordinates. Its charge and magnetic moment distribution are given in terms of charge distribution $\rho(r)$ and 
magnetic moment distributions $\mu(r)$.
Zemach calculate the correction in first order to the hyperfine energy of s-states of hydrogen due to the electric charge distribution. The HFS energy is  written as
\begin{equation}
  \label{eq:nr-hfs}
  \Delta E_{\textrm{HFS}}^{\mathrm{Z}}  = -\frac{2}{3} \left< \boldsymbol{S}_p\cdot\boldsymbol{S}_{\mu}\right> \int \mid \phi(\boldsymbol{r}) \mid^2 \mu(\boldsymbol{r})d \boldsymbol{r} 
\end{equation}
 $\phi$ the non-relativistic electron wavefunction and $S_x$ are the spin operators of the electron and proton. If the magnetic moment distribution is taken to be the one of a point charge,
$\mu(\boldsymbol{r})=\delta (\boldsymbol{r})$, the integral reduces to $ \mid \phi(0) \mid^2$.
The first order correction to the wavefunction due to the nucleus finite charge distribution is given by
\begin{equation}
  \label{eq:wf-first-order}
   \phi(\boldsymbol{r}) =  \phi_{\textrm{C}}(0) \left(1 - \alpha m_{\mu} \int \rho(\boldsymbol{u}) \left | \boldsymbol{u}-\boldsymbol{r}\right | d\boldsymbol{u} \right),
\end{equation}
where $\phi_{\textrm{C}}(0)$ is the unperturbed Coulomb wavefunction at the origin for a point nucleus.
Replacing into Eq.~\eqref{eq:nr-hfs} and keeping only first order terms, we get (Eq.~2.8 of Ref.~\cite{zem1956} corrected for a misprint):
\begin{eqnarray}
  \label{eq:nr-hfs-corr}
  \Delta E_{\textrm{HFS}}^{\mathrm{Z}}  &=& -\frac{2}{3} \left< \boldsymbol{S}_p\cdot\boldsymbol{S}_{\mu}\right>\left|  \phi_{\textrm{C}}(0) \right|^2 \nonumber \\
  && \times \left(1 - 2 \alpha m_{\mu} \int \rho(\boldsymbol{u}) \mid \boldsymbol{u}-\boldsymbol{r} \mid \mu(\boldsymbol{r})d\boldsymbol{u} d\boldsymbol{r} \right),  \nonumber \\
                   &=& E_{\textrm{F}}\left(1 - 2 \alpha m_{\mu} \int \rho(\boldsymbol{u}) \mid \boldsymbol{u}-\boldsymbol{r} \mid \mu(\boldsymbol{r})d\boldsymbol{u} d\boldsymbol{r} \right), \nonumber \\
\end{eqnarray}
where $E_{\textrm{F}}$ is the well known HFS Fermi energy. Transforming Eq.~\eqref{eq:nr-hfs-corr} using $\boldsymbol{r}\to\boldsymbol{r}+\boldsymbol{u}$, etc. Zemach obtains
\begin{equation}
  \label{eq:nr-hfs-zem}
  \Delta E_{\textrm{HFS}}^{\mathrm{Z}}  = E_{\textrm{F}}\left(1 - 2 \alpha m_{\mu} \left<r_{\textbf{Z}}\right>\right),
\end{equation}
with $ \left<r_{\textbf{Z}}\right>$ given in Eq. \eqref{eq:zem-rad}. The $2s$ state Fermi energy is given by
\begin{equation}
\label{eq:fermi-hfs}
E_{\textrm{F}}^{2s}=\frac{(Z\alpha)^4}{3}g_p \frac{\mu_r^3}{m_p m_{\mu}}.
\end{equation}

\subsection{Hyperfine structure of the $2s$ level}
\label{subsec:2shfs}
\begin{figure}[tb]
	\centering
%====================
\includegraphics[width=\columnwidth]{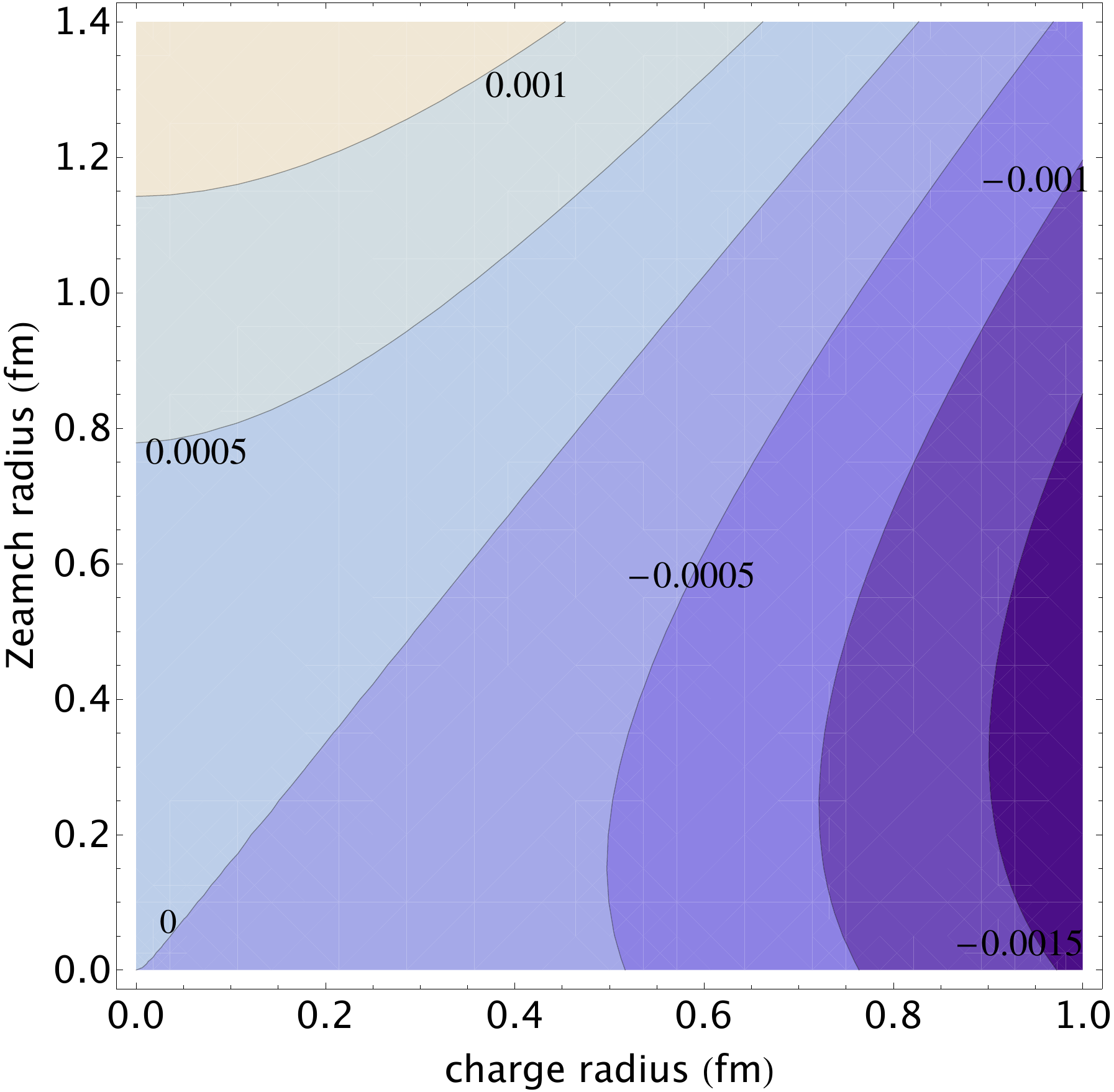}
	\caption[]{difference between Borie's value and Eq. \eqref{eq:fit-2D} result as a function of the charge and  Zemach radii (meV).} \label{fig:borie-vs-pi}
\end{figure}

In order to check the dependence of the hyperfine structure on the Zemach radius and on the proton finite size, I have performed a series of calculations for a dipolar distribution for both the charge and magnetic moment distribution.
We can then study the dependence of the HFS beyond the first order corresponding to the Zemach correction. I calculated the hyperfine energy splitting $\Delta E_{\textrm{HFS}}\left(R_{\textrm{Z}},R\right)=E_{\textrm{HFS}}(R)+ E^{\textrm{BW}}_{\textrm{HFS}}\left(R,R_{M}\right)$ numerically.
 I also evaluate with and without self-consistent inclusion of the Uëhling potential in the calculation, to obtain all-order Uëhling  contribution to the HFS energy.  We calculated the correction  $\Delta E_{\textrm{HFS}}\left(R_{\textrm{Z}},R\right)$ for several value of
  $R_{\textrm{Z}}$ between \unit{0.8}{fm} and \unit{1.15}{fm}, and proton sizes ranging from  \unit{0.3}{fm} to \unit{1.2}{fm}, by steps of \unit{0.05}{fm}, which represents 285 values. The results show that the correction to the HFS energy due to charge and magnetic moment distribution is not quite independent of $R$ as one would expect from
   Eq. \eqref{eq:nr-hfs-zem}, in which the finite size contribution depends \emph{only} on $R_{\textrm{Z}}$. We  fitted  the hyperfine structure splitting of the $2s$ level, $E_{\textrm{HFS}}^{2s}\left(R_{\textrm{Z}},R\right)$ by a function of $R$ and $R_{\textrm{Z}}$, which gives:
\begin{equation}
\label{eq:fit-2D}
\begin{split}
E_{\textrm{HFS}}^{2s}\left(R_{\textrm{Z}},R\right)&=22.807995 \\
& - 0.0022324349 R^2 + 0.00072910794 R^3 \\
&- 0.000065912957 R^4 - 0.16034434 R_{\textrm{Z}} \\
& - 0.00057179529 R R_{\textrm{Z}} \\
& -  0.00069518048 R^2 R_{\textrm{Z}} \\
&- 0.00018463878 R^3 R_{\textrm{Z}} \\
&+ 0.0010566454 R_{\textrm{Z}}^2 \\
& + 0.00096830453 R R_{\textrm{Z}}^2 \\
&+ 0.00037883473 R^2 R_{\textrm{Z}}^2\\
& - 0.00048210961 R_{\textrm{Z}}^3 \\
& -  0.00041573690 R R_{\textrm{Z}}^3\\
& + 0.00018238754 R_{\textrm{Z}}^4\,\textrm{meV.}
\end{split}
%R_{\textrm{Z}}
\end{equation}
The constant term should be close to the sum of the Fermi energy \unit{22.80541}{meV} and of the Breit term \cite{bre1930}. the HFS correction calculated with a point-nucleus Dirac wavefunction for which I find \unit{22.807995}{meV}. When setting the speed of light to infinity in the program I recover exactly the Fermi energy. The Breit contribution is thus \unit{0.002595}{meV}, to be compared to  \unit{0.0026}{meV} in Ref. \cite{mar2005} (Table II, line 3) and \unit{0.00258}{meV} in Ref. \cite{bor2012}.
Martynenko \cite{mar2005} evaluates this correction, which he names ``Proton structure corrections of order $\alpha^5$ and $\alpha^6$'', to be \unit{-0.1535}{meV}, following \cite{pac1996}. He finds the coefficient for the Zemach's radius to be \unit{ -0.16018}{meVf^{-1}}, in very good agreement with the present all-order calculation  \unit{ - 0.16034}{meVf^{-1}}. Borie's value \cite{bor2012}  \unit{ -0.16037}{meVf^{-1}} is even closer. The difference between Borie's value and Eq. \eqref{eq:fit-2D} is represented in Fig. \ref{fig:borie-vs-pi} as a function of the charge and Zemach radii. The maximum difference is around \unit{1}{\mu eV}.

 In Ref. \cite{pac1996}, the charge and magnetic moment distributions are written down in the dipole form, which corresponds to \eqref{eq:exp-formfac},
\begin{equation}
\label{eq:pachu-dipole}
G_{E}\left(q^2\right)=\frac{G_{M}\left(q^2\right)}{1+\kappa_p}=\frac{\Lambda^4}{\left(\Lambda^2+q^2\right)^2},
\end{equation}
with $\Lambda=848.5$MeV. This leads to $R=0.806$~fm as in Ref. \cite{ssbw1980} and $R_{\textrm{Z}}=1.017$~fm using this definition for the form factor in Eq. \eqref{eq:zemq}. Moreover there are recoil corrections included. Pachucki \cite{pac1996} finds that the pure Zemach contribution (in the limit $m_p\to\infty$) is \unit{-0.183}{meV}. In Ref. \cite{cfm2002}, the Zemach corrections is given as $\delta(\textrm{Zemach})\times E_{F}=-71.80\times 10^{-4}E_F$, for $R_{\textrm{Z}}=1.022$~fm. This leads to a coefficient \unit{-0.1602}{meVfm}$^{-1}$, in excellent agreement with our value \unit{-0.16036}{meVfm}$^{-1}$.

The effect of the vacuum polarization on the $2s$ hyperfine structure energy shift as a function of the Zemach and charge radius have been calculated for the same set of values as  the main contribution.
% A set of value can be found in the last column of Table  \ref{tab:2shfs} for  $R_{\textrm{Z}}=1.045$fm. 
The data can be described as a function of $R_{\textrm{Z}}$ and $R$ as
%\begin{equation}
%\label{eq:2s-HFS-VP-2D-fit}
%\begin{split}
%E_{\textrm{HFS}}^{2s,VP}\left(R_{\textrm{Z}},R\right)&=-0.000025339 R^2+0.000154707 R_{\textrm{Z}}^2 \\
% & -0.00203434R_{\textrm{Z}}+0.0744207 \textrm{meV.}
%\end{split}
%\end{equation}
%New:
\begin{equation}
\label{eq:2s-HFS-VP-2D-fit}
\begin{split}
E_{\textrm{HFS}}^{2s,VP}\left(R_{\textrm{Z}},R\right)&=
%0.0744057  \\
%&+ 0.00001045172 R^2 \\
%&- 0.00200252 R_{\textrm{Z}} \\
%& - 0.0000556776 R R_{\textrm{Z}}\\
% &+ 0.000157724 R_{\textrm{Z}}^2
% updated July 22nd, 2012
0.074369030 + 0.000074236132 R^2 \\
&  + 0.00013277334 R^3 -  8.0987285\times10^{-6} R^4 \\
&  - 0.0017880269 R_{\textrm{Z}} - 0.00017204505 R R_{\textrm{Z}} \\
&  -  0.00037499458 R^2 R_{\textrm{Z}} \\
& - 0.000070355379 R^3 R_{\textrm{Z}} \\
&  - 0.00022093411 R_{\textrm{Z}}^2 +  0.00035038656 R R_{\textrm{Z}}^2 \\
& + 0.00020554316 R^2 R_{\textrm{Z}}^2 + 0.00025100642 R_{\textrm{Z}}^3 \\
&  - 0.00017200435 R R_{\textrm{Z}}^3\\
& - 0.000061266973 R_{\textrm{Z}}^4 \,\textrm{meV.}
\end{split}
\end{equation}

\begin{figure}[tb]
	\centering
%====================
\includegraphics[width=\columnwidth]{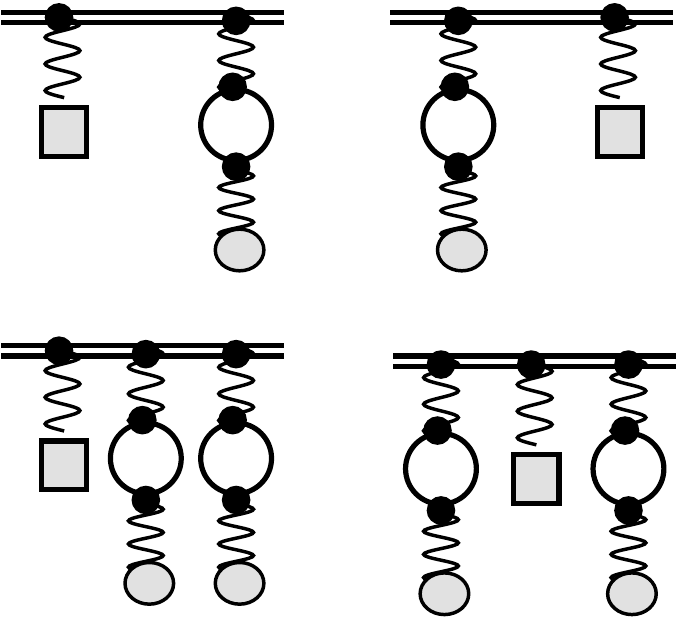}
	\caption[]{Feynman diagrams corresponding to the evaluation of the hyperfine structure using  wavefunctions obtained with the Uehling potential in the Dirac equation. The grey squares correspond to the hyperfine interaction. } \label{fig:hfs-vp-diag}
\end{figure}

It corresponds to the diagrams presented in Fig. \ref{fig:hfs-vp-diag}. The size-independent term \unit{0.07437}{meV}  corresponds to the sum of
the two contributions  represented by the two top diagrams in Fig. \ref{fig:hfs-vp-diag} and is given as  \unit{\Delta E^{HFS}_{1 \textrm{loop-after-loop VP}}=0.0746}{meV} in Ref. \cite{mar2005}. The term $\Delta E^{HFS}_{1\gamma, \textrm{VP}}=0.0481$~meV corresponds to a vacuum polarization loop in the HFS potential \cite{bar1982,pac1996,mar2005}, which is not evaluated here.
Corrections present in Ref. \cite{bor2012} not included in Eqs.   \eqref{eq:fit-2D} and \eqref{eq:2s-HFS-VP-2D-fit} gives an extra contribution of 
\begin{equation}
\label{eq:2s-HFS-HO}
E_{\textrm{HFS}}^{2s,HO}=0.10287  \,\textrm{meV.}
\end{equation}

Combining Eqs. \eqref{eq:fit-2D}, \eqref{eq:2s-HFS-VP-2D-fit} and \eqref{eq:2s-HFS-HO}, I get
\begin{equation}
\label{eq:2s-HFS-final}
\begin{split}
E_{\textrm{HFS}}^{2s}\left(R_{\textrm{Z}},R\right)
%&=22.972964 \\
%&- 0.0021581988 R^2 \\
%&+ 0.00086188128 R^3 \\
% &-  0.000074011685 R^4\\
% & - 0.16213237 R_{\textrm{Z}} \\
% &- 0.00074384033 R R_{\textrm{Z}} \\
% & -  0.0010701751 R^2 R_{\textrm{Z}} \\
% &- 0.00025499415 R^3 R_{\textrm{Z}} \\
% & + 0.00083571133 R_{\textrm{Z}}^2 \\
%& + 0.0013186911 R R_{\textrm{Z}}^2\\
%& + 0.00058437789 R^2 R_{\textrm{Z}}^2 \\
%&- 0.00023110319 R_{\textrm{Z}}^3 \\
%&-  0.00058774125 R R_{\textrm{Z}}^3 \\
%&+ 0.00012112057 R_{\textrm{Z}}^4 \\
%updated July 22nd, 2012
&=22.985234 - 0.0021581988 R^2 \\
& + 0.00086188128 R^3 \\
&- 0.000074011685 R^4 \\
&  - 0.16213237 R_{\textrm{Z}} \\
& - 0.00074384033 R R_{\textrm{Z}} \\
& -  0.0010701751 R^2 R_{\textrm{Z}} \\
& - 0.00025499415 R^3 R_{\textrm{Z}} \\
& + 0.00083571133 R_{\textrm{Z}}^2 \\
& + 0.0013186911 R R_{\textrm{Z}}^2 \\
& + 0.00058437789 R^2 R_{\textrm{Z}}^2 \\
& - 0.00023110319 R_{\textrm{Z}}^3 \\
& -  0.00058774125 R R_{\textrm{Z}}^3 \\
& + 0.00012112057 R_{\textrm{Z}}^4 
  \,\textrm{meV.}
\end{split}
\end{equation}
In Ref. \cite{mar2005}, the equivalent expression is
\begin{equation}
\label{eq:2s-HFS-Mart}
E_{\textrm{HFS}}^{2s\textrm{Mart.}}\left(R_{\textrm{Z}}\right)=22.9857 - 0.16018 R_{\textrm{Z}}
 \,\textrm{meV},
\end{equation}
while it is
\begin{equation}
\label{eq:2s-HFS-Borie}
E_{\textrm{HFS}}^{2s\textrm{Borie}}\left(R_{\textrm{Z}}\right)=22.9627-0.16037 R_{\textrm{Z}}
 \,\textrm{meV},
\end{equation}
in Ref. \cite{bor2012}.
Using a Zemach's radius of \unit{0.9477}{fm} in Eq. \eqref{eq:2s-HFS-Mart}, needed to reproduce entry 11 in Table II of Ref. \cite{mar2005}, one obtains \unit{22.8148}{meV} as expected. In Eq. \eqref{eq:2s-HFS-Borie}, it gives \unit{22.8107}{meV}. Using the same Zemach radius and Eq. \eqref{eq:2s-HFS-final} I obtain
 \unit{22.8104}{meV} with the muonic hydrogen proton radius value and  \unit{22.8103}{meV} with the CODATA one, in excellent agreement with Borie's value. All three values are in agreement with the result \unit{22.8146(49)}{meV} in Ref. \cite{cng2011}.
 In a recent work, however, the use of Form factors in the Breit equations leads to  smaller finite size corrections, leading to \unit{22.8560}{meV} \cite{dkn2012}. A comparison between some of these results is presented in Table \ref{tab:2shfs}.

\begin{table*}[tbp]
\caption{Comparison of contributions to the $2s$ hyperfine structure from Refs. \cite{mar2005,bor2012} and the present work (meV) for  \unit{R_{\textrm{Z}}=1.0668}{fm} \cite{ssbw1980} as used in  Ref. \cite{mar2005}. Note that in this reference,  the proton structure correction of order  $\alpha^5$ (item \# 6) may combine the Zemach correction and the recoil correction (\# 24). VP: Vacuum Polarization.
}
\begin{center}
\squeezetable
\begin{ruledtabular}
\begin{tabular}{lcddd}
	&\#&\multicolumn{1}{c}{Ref. \cite{mar2005}}	&	\multicolumn{1}{c}{Ref. \cite{bor2012}}	&	\multicolumn{1}{c}{This work}	\\
	\hline
Fermi energy	&	1	&	22.8054	&	22.8054	&		\\
Dirac Energy (includes Breit corr.)	&	2	&		&		&	22.807995	\\
Vacuum polarization corrections of orders $\alpha^5, \alpha^6$
in  2nd-order 	&	3	&	0.0746	&	0.07443	&		\\
perturbation theory $\epsilon_{VP1}$&&&\\
All-order VP contribution to HFS, with finite magnetisation distribution 	&	4	&		&		&	0.07244	\\
finite extent of magnetisation density correction to the above	&	5	&		&	-0.00114	&		\\
Proton structure corr. of order  $\alpha^5$	&	6	&	-0.1518	&	-0.17108	&	-0.17173	\\
Proton structure corrections of order $\alpha^6$	&	7	&	-0.0017	&		&		\\
Electron vacuum polarization contribution+
proton structure corrections of order $\alpha^6$	&	8	&	-0.0026	&		&		\\
contribution of $1\gamma$ interaction of order $\alpha^6$	&	9	&	0.0003	&	0.00037	&	0.00037	\\
$\epsilon_{VP} 2 E_F$ (neglected in Ref. \cite{mar2005})	&	10	&		&	0.00056	&	0.00056	\\
muon loop VP  (part corresponding to $\epsilon_{VP2}$ neglected in Ref. \cite{mar2005})	&	11	&		&	0.00091	&	0.00091	\\
Hadronic Vac. Pol.	&	12	&	0.0005	&	0.0006	&	0.0006	\\
Vertex (order $\alpha^5$)	&	13	&		&	-0.00311	&	-0.00311	\\
Vertex (order $\alpha^6$) (only part with powers of $\ln(\alpha)$ - see Ref. \cite{bae1966} )	&	14	&		&	-0.00017	&	-0.00017	\\
Breit	&	15	&	0.0026	&	0.00258	&		\\
Muon anomalous magnetic moment correction of order $\alpha^5$, $\alpha^6$	&	16	&	0.0266	&	0.02659	&	0.02659	\\
Relativistic and radiative recoil corrections
with  &	17	&	0.0018	&		&		\\
proton anomalous magnetic moment of order  $\alpha^6$	&&&\\
One-loop electron vacuum polarization
contribution of  $1\gamma$ interaction&	18	&	0.0482	&	0.04818	&	0.04818	\\
 of orders $\alpha^5$, $\alpha^6$ ($\epsilon_{VP2}$)	 &&&\\
finite extent of magnetisation density correction to the above	&	19	&		&	-0.00114	&	-0.00114	\\
One-loop muon vacuum polarization
contribution of $1\gamma$ interaction of order  $\alpha^6$	&	20	&	0.0004	&	0.00037	&	0.00037	\\
Muon self energy+proton structure
correction of order $\alpha^6$	&	21	&	0.001	&		&	0.001	\\
Vertex corrections+proton structure
corrections of order $\alpha^6$	&	22	&	-0.0018	&		&	-0.0018	\\
``Jellyfish'' diagram correction+
proton structure corrections of order $\alpha^6$	&	23	&	0.0005	&		&	0.0005	\\
Recoil correction Ref. \cite{cng2008}	&	24	&		&	0.02123	&	0.02123	\\
Proton polarizability contribution of order  $\alpha^5$	&	25	&	0.0105	&		&		\\
Proton polarizability  Ref. \cite{cng2008}	&	26	&		&	0.00801	&	0.00801	\\
Weak interaction contribution	&	27	&	0.0003	&	0.00027	&	0.00027	\\
\hline									
Total	&		&	22.8148	&	22.8129	&	22.8111	\\
		\end{tabular}
\end{ruledtabular}
\label{tab:2shfs}
\end{center}
\end{table*}

%A summary of all known contributions from previous work and the present one as presented in Table \ref{tab:size-ind-2s-HFS} for the part independent of charge and magnetic moment distribution, and in Table for the part depending on the charge and Zemach's radiii.

%\begin{figure}[tb]
%	\centering
%%====================
%\includegraphics[width=\columnwidth]{fit-2D-HFS-2s-RZ}
%	\caption[]{Finite charge and magnetic moment distribution energy shift for the $2s$ state as a function of the charge $R$ and Zemach radius $R_{\textrm{Z}}$ for the Gaussian ( $R_{\textrm{Z}}=1.045$fm) and exponential model, divided by $R_{\textrm{Z}}$. The lines correspond to the function in Eq. \eqref{eq:fit-2D}. } \label{fig:2shfs}
%\end{figure}
%
%
%\begin{figure}[tb]
%	\centering
%%====================
%\includegraphics[width=\columnwidth]{HFS-vp-corr-2s-R-R_{\textrm{Z}}}
%	\caption[]{Effect of vacuum polarization  on the $2s$ state as a function of the charge $R$ and Zemach radius $R_{\textrm{Z}}$ for the Gaussian ( $R_{\textrm{Z}}=1.045$fm) and exponential model. The lines correspond to the function in Eq. \eqref{eq:2s-HFS-VP-2D-fit}. } \label{fig:2shfs-vp}
%\end{figure}

\section{Evaluation of muonic hydrogen $n=2$ transitions}
\label{sec:final-val-trans}
\subsection{Lamb shift and fine structure}
\label{subsec:lamb-fs}
The results presented in this work for the Lamb shift (Eqs. \eqref{eq:coulfsr6log2}, \eqref{eq:vp11cfs}, \eqref{eq:ks21fs}, \eqref{eq:itervp}, \eqref{eq:vp11xks}, \eqref{eq:wk}, \eqref{eq:muon-LS-se-fs}, \eqref{eq:muonvpfs}, \eqref{eq:muonvpfsfs}) can be summarized in the following proton-size dependent equation:
\begin{equation}
\begin{split}
\Delta E^{\mathrm{Tot, fs}}_{2s_{1/2}-2p_{1/2}}(R)&=
% 206.020925\\
% & - 5.2278228 R^2 \\
% &+ 0.036999 R^3 \\
% & - 0.00116103 R^4 \\
% & + 0.00044095 R^5 \\
% & -  0.00006662 R^6 
% new version, May 21st 2012
% updated 21 july 2012
206.0209137 - 5.226135625 R^2 \\
& + 0.03432100160 R^3 \\
& + 0.0005454642475 R^4 \\
&-0.00008785574420R^5\\
& + 0.0002962967640 R^2 \log(R) \\
& - 0.00004751147090 R^4 \log(R)
 \,\textrm{meV}.
\end{split}
\label{eq:tot-all-order-LS}
\end{equation}
In the same way, Eqs. \eqref{eq:coulfsfsc},  \eqref{eq:vp11fsfs}, \eqref{eq:ks21fsfs}, \eqref{eq:itervpfs}, \eqref{eq:vp11xksfsfs},  \eqref{eq:wkfs}, \eqref{eq:muon-FS-se-fs},  \eqref{eq:muonvpfsfs-fs}, and \eqref{eq:recoil-2p1-2p3} lead to the fine structure interval (which include the recoil corrections \eqref{eq:recoil-2p1-2p3}, included in Table \ref{tab:other-contribs-LS} for the $2s$ Lamb shift)
\begin{equation}
\begin{split}
\Delta  E^{\mathrm{Tot, fs}}_{2p_{1/2}-2p_{3/2}}(R) &=
%&=8.438311 \\
%&  - 0.000053037413 R^2 \\
%&   + 9.33506\times 10^{-7} R^3 \\
%&    - 4.33656\times 10^{-7} R^4 \\
%&    +  1.2769945\times 10^{-7} R^5\\
%&      - 1.5604806\times 10^{-8} R^6 \\
%
%&=  8.3520516 \\
%& - 0.00005206070 R^2 \\
%&+ 1.70858\times10^{-7} R^3 \\
%&-  4.30422\times10^{-8} R^4 \\
%&+ 1.54724\times10^{-8} R^5\\
%& - 2.13593\times10^{-9} R^6
% Udated october 29, 2012
%8.420695454 - 0.00005203798087 R^2 + 1.215060759*10^-7 R^3 - 
 %1.544056441*10^-9 R^4
%&=  8.420695454 \\
%& - 0.00005203798087  R^2 \\
%&+ 1.215060759\times10^{-7} R^3 \\
%&-  1.544056441\times10^{-9} R^4 
% updated Noc. 1st, 2012
8.352051651\\
& - 0.00005203798087 R^2 \\
&+ 1.215060759 \times 10^{-7} R^3\\
& - 1.544056441 \times 10^{-9} R^4
  \,\textrm{meV}.
\end{split}
\label{eq:tot-all-order-FS}
\end{equation}
Martynenko \cite{mar2008} finds \unit{E^{\mathrm{Tot, fs}}_{2p_{1/2}-2p_{3/2}}=8.352082}{meV} for the fine structure.

A number of terms not included in Eqs. \eqref{eq:tot-all-order-LS}  are presented in Table \ref{tab:other-contribs-LS} together with the relevant references.
Combining Eq. \eqref{eq:tot-all-order-LS} with the sum of the contributions contained in Table \ref{tab:other-contribs-LS}, I obtain the final  $2s-2p_{1/2}$ energy:
\begin{equation}
\begin{split}
\Delta  E^{\mathrm{Tot, fs}}_{2s_{1/2}-2p_{1/2}}(R)&=
% 206.063525482\\
% &-5.2278228 R^2 \\
% &+ 0.036998584 R^3\\
% &- 0.0011610273 R^4 \\
% &+ 0.00044095136 R^5 \\
% &- 0.000066621019 R^6 
% new version, June 3rd 2012
% 206.0617141 - 5.227821393 R^2 \\
%& + 0.03533582447 R^3 + 
% 0.00005581440961 R^4\\
% & + 4.750603833\times10^{-6} R^5 \\
% & - 5.247051060\times10^{-7} R^6 \\
% & + 0.0002946564258 R^2 \log(R) \\
% & - 0.00004957597920 R^4 \log(R)
%new version July 21st 2012
%update October 27
206.0465137 - 5.226135625 R^2 \\
& + 0.03432100160 R^3 \\
& +  0.0005454642475 R^4  \\
&-0.00008785574420R^5\\
 & + 0.0002962967640R^2 \log(R)\\
 & - 0.00004751147090 R^4 \log(R)
 \,\textrm{meV}.
\end{split}
\label{eq:Final-all-order-LS}
\end{equation}
This can be compared with the result from  from E.\ Borie \cite{bor2012}
\begin{equation}
\label{eq:borie-res}
 \begin{split}
\Delta E^{\mathrm{Borie, fs}}_{2s_{1/2}-2p_{1/2}}(R)&=206.0579(60)-5.22713 R^2  \\
&+ 0.0365(18)R^3\,\textrm{meV},
 \end{split}
\end{equation}
and Carroll et al. \cite{ctrm2011}
\begin{equation}
\label{eq:carroll-res}
 \begin{split}
\Delta  E^{\mathrm{Carlson fs}}_{2s_{1/2}-2p_{1/2}}(R)&=206.0604-5.2794 R^2  \\
&+ 0.0546R^3\,\textrm{meV}.
\end{split}
\end{equation}

An extra recoil contribution is given in Ref. \cite{kik2012} for the fine structure, corresponding to entry \#9  in  Table \ref{tab:other-contribs-LS} for the Lamb shift,
\begin{equation}
\Delta  E^{\mathrm{VP Rec.}}_{2p_{1/2}-2p_{3/2}}=-0.00006359  \,\textrm{meV}.
\end{equation}
This term correspond to corrections beyond the full Dirac term. This lead to the final result
\begin{equation}
\begin{split}
\Delta  E^{\mathrm{Tot, fs}}_{2p_{1/2}-2p_{3/2}}(R) &=
%&=8.438311 \\
%&  - 0.000053037413 R^2 \\
%&   + 9.33506\times 10^{-7} R^3 \\
%&    - 4.33656\times 10^{-7} R^4 \\
%&    +  1.2769945\times 10^{-7} R^5\\
%&      - 1.5604806\times 10^{-8} R^6 \\
%
%&=  8.357015415 \\
%& - 0.00005206070 R^2 \\
%&+ 1.70858\times10^{-7} R^3 \\
%&-  4.30422\times10^{-8} R^4 \\
%&+ 1.54724\times10^{-8} R^5\\
%& - 2.13593\times10^{-9} R^6
% July 21, 2012
%&=8.351988025 \\
%& - 0.00005206070023 R^2 \\
%& + 1.708581114\times 10^{-7} R^3 \\
%&-  4.304222172\times 10^{-8} R^4  \\
%&+ 1.547238809\times 10^{-8} R^5 \\
%&- 2.135927044\times 10^{-9} R^6
% updated October 29, 2012
8.351988061 \\
&- 0.00005203798087 R^2 \\
&+ 1.215060759 \times 10^{-7} R^3 \\
& - 1.544056441 \times 10^{-9} R^4
   \,\textrm{meV}.
\end{split}
\label{eq:tot-all-order-FS-VP}
\end{equation}

\begin{table*}[tbp]
\caption{Contributions to the Lamb shift not included in Eq. \eqref{eq:tot-all-order-LS} (meV). The uncertainty on the proton polarization value used in Ref. \cite{pana2010} has been increased by a factor of 10, according to the discussion in Ref. \cite{hap2011a}.}
\squeezetable
\begin{center}
\begin{ruledtabular}
\begin{tabular}{clcccdd}
\#	&	Contribution	&	\multicolumn{1}{c}{Reference}	&	\multicolumn{1}{c}{Value}	&	\multicolumn{1}{c}{Unc.}	\\
\hline									
1	&	NR three-loop electron VP (Eq. (11), (15), (18) and (23))	&	\cite{kan1999}	&	0.00529	&		\\
2	&	Virtual Delbrück scattering (2:2)	&	\cite{kiks2010,kkis2010}	&	0.00115	&	0.00001	\\
3	&	Light by light electron loop contribution (3:1)	&	\cite{kiks2010,kkis2010}	&	-0.00102	&	0.00001	\\
\hline									
4	&	Mixed self-energy vacuum polarization	&	\cite{pac1996,jen2011,jaw2011} 	&	-0.00254	&		\\
5	&	Hadronic  vacuum polarization	&	\cite{fms1999,maf2000,maf2001}	&	0.01121	&	0.00044	\\
\hline									
6	&	Recoil contribution Eqs. \eqref{eq:rec1} and \eqref{eq:rec2}	&	\cite{bag1955,say1990,vap2004,mtn2008}	&	0.05747063	&		\\
7	&	Relativistic recoil of order $(Z\alpha)^5$ Eq. \eqref{eq:rr1}	&	\cite{pac1999,egs2001,egs2007,bor2005a,mtn2008}	&	-0.04497053	&		\\
8	&	Relativistic Recoil of order $(Z\alpha)^6$ Eq. \eqref{eq:rr2s}	&	\cite{pac1999,mtn2008}	&	0.0002475	&		\\
9	&	Recoil correction to VP of order $m/M$ and $(m/M)^2$  in Eq. (4)	&	\cite{kik2012}	&	-0.001987	&		\\
\hline									
10	&	Proton Self-energy	&	\cite{pac1996,pac1999,egs2007,hap2011}	&	-0.0108	&	0.0010	\\
11	&	Proton polarization	&	\cite{pac1999,ros1999,maf2000,mar2006,cav2011}	&	0.0129	&	0.0040	\\
\hline									
12	&	Electron loop in the radiative photon	&	\cite{saw1957,pet1957,bcr1973,bar1982}	&	-0.00171	&		\\
	&	of order  $\alpha^2(Z\alpha)^4$	&		&		&		\\
13	&	Mixed electron and muon loops	&	\cite{bor1975}	&	0.00007	&		\\
14	&	Rad. Recoil corr. $\alpha(Z\alpha)^5$	&	\cite{jen2011a}	&	0.000136	&		\\
15	&	Hadronic polarization $\alpha(Z\alpha)^5m_r$	&	\cite{maf2000,maf2001}	&	0.000047	&		\\
16	&	Hadronic polarization in the radiative 	&	\cite{maf2000,maf2001}	&	-0.000015	&		\\
	&	photon $\alpha^2(Z\alpha)^4m_r$	&		&		&		\\
17	&	Polarization operator induced correction	&	\cite{maf2001}	&	0.00019	&		\\
	&	 to nuclear polarizability $\alpha(Z\alpha)^5m_r$	&		&		&		\\
18	&	Radiative photon induced correction 	&	\cite{maf2001}	&	-0.00001	&		\\
	&	to nuclear polarizability $\alpha(Z\alpha)^5m_r$	&		&		&		\\
\hline									
	&	Total	&		&	0.0256	&	0.0041	\\
\end{tabular}
\end{ruledtabular}
\label{tab:other-contribs-LS}
\end{center}
\end{table*}

\subsection{Transitions between hyperfine sublevels}
\label{subsec:hfs-trans}
The energies of the two transitions observed experimentally in muonic hydrogen are given by
\begin{equation}
\begin{split}
\label{eq:hfs-F2-F1}
E^{F=2}_{2p_{3/2}}- E^{F=1}_{2s_{1/2}}&=\Delta E_{2s_{1/2}-2p_{1/2}}
 +\Delta E_{2p_{1/2}-2p_{3/2}} \\ 
& +\frac{3}{8}E_{\textrm{HFS}}^{2p_{3/2}}  - \frac{1}{4}E_{\textrm{HFS}}^{2s} \, ,
 \end{split}
\end{equation}
and
\begin{equation}
\begin{split}
\label{eq:hfs-F1-F0}
E^{F=1}_{2p_{3/2}}- E^{F=0}_{2s_{1/2}}&=\Delta E_{2s_{1/2}-2p_{1/2}}
 +\Delta E_{2p_{1/2}-2p_{3/2}} \\
 & -\frac{5}{8}E_{\textrm{HFS}}^{2p_{3/2}} + \frac{3}{4}E_{\textrm{HFS}}^{2s} + \delta E_{\textrm{HFS}}^{F=1} \, .
 \end{split}
\end{equation}
Here we use the results from \cite{mar2008} for the $2p$ states:
\begin{equation}
\begin{split}
\label{eq:hfs-2p-mart}
E_{\textrm{HFS}}^{2p_{1/2}} &= 7.964364 \,\textrm{meV}\\
E_{\textrm{HFS}}^{2p_{3/2}} &= 3.392588 \,\textrm{meV} \\
\delta E_{\textrm{HFS}}^{F=1}&=  0.14456\,\textrm{meV}.
 \end{split}
\end{equation}

Using the results presented above I get
\begin{equation}
\begin{split}
\label{eq:hfs-F2-F1-final}
E^{F=2}_{2p_{3/2}}\left(R_{\textrm{Z}},R\right)- E^{F=1}_{2s_{1/2}}\left(R_{\textrm{Z}},R\right)&=
209.92441 \\
&- 5.2261075 R^2\\
& + 0.034379506 R^3 \\
&+ 0.00043287446 R^4 \\
&- 
 0.000063788419 R^5 \\
 &+ 0.040533092 R_{\textrm{Z}} \\
 &+ 0.00018596008 R R_{\textrm{Z}}\\
 & + 
 0.00026754376 R^2 R_{\textrm{Z}} \\
 &+ 0.000063748539 R^3 R_{\textrm{Z}} \\
 & - 0.00020892783 R_{\textrm{Z}}^2 \\
 &- 
 0.00032967277 R R_{\textrm{Z}}^2 \\
 &- 0.00014609447 R^2 R_{\textrm{Z}}^2\\
 & + 
 0.000057775798 R_{\textrm{Z}}^3 \\
 & + 0.00014693531 R R_{\textrm{Z}}^3 \\
 &- 0.000030280142 R_{\textrm{Z}}^4 \\
 &+ 
 0.00029629676 R^2 \log(R) \\
 &- 0.000047511471 R^4 \log(R)
& \qquad\qquad \,\textrm{meV.}
 \end{split}
\end{equation}

This can be compared with the result from U.~Jentschura \cite{jen2011}
\begin{equation}
\label{eq:jentsc-res}
 \begin{split}
 E^{\mathrm{Jents.},F=2}_{2p_{3/2}}- E^{\mathrm{Jents.},F=1}_{2s_{1/2}}&=209.9974(48)\\
& -5.2262 R^2 \,\textrm{meV},
 \end{split}
\end{equation}
using the $2s$ hyperfine structure of Ref. \cite{mar2005}.

For the other transition I obtain
\begin{equation}
\begin{split}
\label{eq:hfs-F1-F0-final}
E^{F=1}_{2p_{3/2}}\left(R_{\textrm{Z}},R\right)- E^{F=0}_{2s_{1/2}}\left(R_{\textrm{Z}},R\right)&=
229.66162 \\
&- 5.2282657 R^2 \\
&+ 0.035241387 R^3 \\
&+ 0.00035886278 R^4\\
& - 
 0.000063788419 R^5 \\
 &- 0.12159928 R_{\textrm{Z}} \\
 &- 0.00055788025 R R_{\textrm{Z}}\\
 & - 
 0.00080263129 R^2 R_{\textrm{Z}} \\
 &- 0.00019124562 R^3 R_{\textrm{Z}} \\
 &+ 0.00062678350 R_{\textrm{Z}}^2 \\
 &+ 
 0.00098901832 R R_{\textrm{Z}}^2 \\
 & + 0.00043828342 R^2 R_{\textrm{Z}}^2 \\
 &- 0.00017332740 R_{\textrm{Z}}^3 \\
 &- 
 0.00044080593 R R_{\textrm{Z}}^3 \\
 &+ 0.000090840426 R_{\textrm{Z}}^4 \\
 &+ 
 0.00029629676 R^2 \log(R)\\
 & - 0.000047511471 R^4 \log(R)
& \qquad\qquad \,\textrm{meV.}
 \end{split}
\end{equation}

\begin{figure}[tb]
	\centering
%====================
\includegraphics[width=\columnwidth]{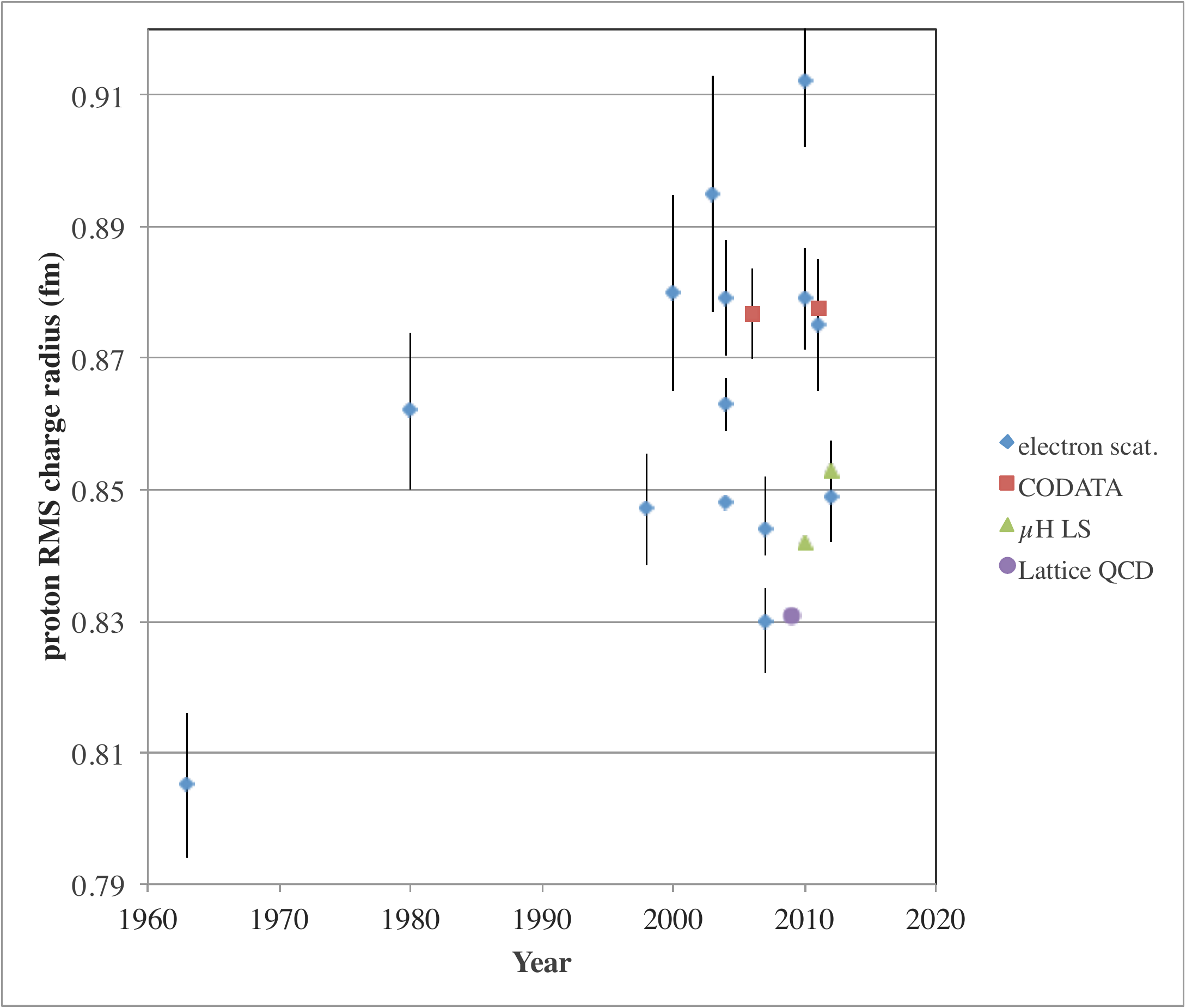}
	\caption[]{Plot of the proton size as a function of time and method. } \label{fig:prot-size-plot}
\end{figure}

Using Eq. \eqref{eq:hfs-F2-F1-final}, a Zemach radius of \unit{1.0668}{fm} from Ref. \cite{mar2005} and the transition energy from Ref. \cite{pana2010}, I obtain a charge radius for the proton of \unit{0.84091(69)}{fm} in place of \unit{0.84184(69)}{fm} in Ref. \cite{pana2010} and \unit{0.8775(51)}{fm} in the 2010 CODATA fundamental constant adjustment. This is \unit{7.1}{\sigma} (using the combined $\sigma$) from the 2010 CODATA value. 
A summary of proton size determinations is presented in Table \ref{tab:prot-size-sum} and Fig. \ref{fig:prot-size-plot}.

\begin{table*}[tbp]
\caption{Proton size determinations (fm). e$^-$p: electron-proton scattering, $\mu$H: muonic hydrogen, ChPt: Lattice QCD corrected with Chiral perturbation theory. The values in the four last lines are obtained using 
the transition frequency from Ref. \cite{pana2010}}
\squeezetable
\begin{center}
\begin{ruledtabular}
\begin{tabular}{ldcdc}
Hand et al. \cite{hmw1963}	&	0.805	&	$\pm$	&	0.011	&	e$^-$p	\\
Simon et al. \cite{ssbw1980}	&	0.862	&	$\pm$	&	0.012	&	e$^-$p	\\
Mergel et al. \cite{mmd1996}	&	0.847	&	$\pm$	&	0.008	&	e$^-$p	\\
Rosenfelder \cite{ros2000}	&	0.880	&	$\pm$	&	0.015	&	e$^-$p	\\
Sick 2003 \cite{sic2003}	&	0.895	&	$\pm$	&	0.018	&	e$^-$p	\\
Angeli \cite{ang2004}	&	0.8791	&	$\pm$	&	0.0088	&	e$^-$p	\\
Kelly \cite{kel2004}	&	0.863	&	$\pm$	&	0.004	&	e$^-$p	\\
Hammer et al. \cite{ham2004}	&	0.848	&		&		&	hydrogen, e$^-$p	\\
CODATA 06 \cite{mtn2008}	&	0.8768	&	$\pm$	&	0.0069	&	Hydrogen, e$^-$p	\\
Arington et al. \cite{amt2007}	&	0.850	&		&		&	e$^-$p	\\
Belushkin et al. \cite{bhm2007} SC approach	&	0.844	&	$^{-0.004}_{+0.008}$	&		&	e$^-$p	\\
Belushkin et al. \cite{bhm2007}pQCD app.	&	0.830	&	$^{-0.008}_{+0.005}$	&		&	e$^-$p	\\
Wang et al. \cite{wlty2009}	&	0.828	&		&		&	ChPt	\\
Pohl et al. \cite{pana2010}	&	0.84184	&	$\pm$	&	0.00067	&	$\mu$H	\\
Bernauer et al. \cite{baab2010}	&	0.879	&	$\pm$	&	0.008	&	e$^-$p	\\
CODATA 2010 \cite{CODATA2010}	&	0.8775	&	$\pm$	&	0.0051	&	Hydrogen, e$^-$p	\\
Adamu\v{s}\v{c}ín et al. \cite{add2011,add2012}	&	0.84894	&	$\pm$	&	0.00690	&	e$^-$p	\\
% updated October 7th, 2012
This work	(using \unit{R_{\textrm{Z}}=1.045}{fm})\cite{vsps2005}&	0.84079	&	$\pm$	&	0.00069	&	$\mu$H	\\
This work	(using \unit{R_{\textrm{Z}}=1.0668}{fm})\cite{mar2005}&	0.84089	&	$\pm$	&	0.00069	&	$\mu$H	\\
Using Jentschura \cite{jen2011}	&	0.84169	&	$\pm$	&	0.00066	&	$\mu$H	\\
Using Borie 	&	0.84232	&	$\pm$	&	0.00069	&	$\mu$H	\\
\end{tabular}
\end{ruledtabular}
\label{tab:prot-size-sum}
\end{center}
\end{table*}

\section{Conclusion}
\label{sec:concl}
In the present work, I have evaluated finite-size dependent contributions to the $n=2$ Lamb shift in muonic hydrogen, to the fine structure and to the $2s$ hyperfine splitting. The calculations were performed numerically, to all order in the finite size correction, in the framework of the Dirac equation. High-order size contributions to the Uelhing potential and to higher-order QED corrections been evaluated. The full dependance of the $2s$  hyperfine splitting on the proton charge distribution and Zemach radius has been evaluated as well.  

The  discrepancy between the proton size deduced from muonic hydrogen and the one coming from CODATA is slightly enlarged  when tacking into account all the newly calculated effects. It is changed from \unit{6.9}{\sigma} to \unit{7.1}{\sigma}.

\acknowledgments
The author wishes to thank Randof Pohl, Eric-Olivier Le Bigot, François Nez, François Biraben  and several other members of the \emph{CREMA} collaboration for numerous and enlightening discussions and Jean-Paul Desclaux for help in implementing some new corrections in the MCDF code. Special thanks go to Peter Mohr for several invitation to NIST where part of this work was performed, many discussions and for critical reading of part of the manuscript. I also thank Michael Distler for providing me with the charge density deduced from the MAMI experiment and Krzysztof Pachuki for his suggestion to introduce the logarithmic contribution to the fits. I thank also Franz Kottmann and Aldo Antognini for a critical and detailed reading of the manuscript.

The Feynman diagrams presented in the figures are realized with JAXODRAW \cite{bckt2009}.
This research was partly supported by the Helmholtz Alliance HA216/EMMI. Laboratoire Kastler Brossel is ``Unité Mixte de Recherche n° 8552'' of École Normale Supérieure, CNRS and Université Pierre et Marie Curie.

\bibliography{muhyd}

\begin{appendix}
\section{Coefficients for the numerical evaluation of the Källén and Sabry potential for a point nucleus}
\label{app:v21pn}
The functions defined in Eq. \eqref{eq:L1shape} are given here. We find, for $x\leq3$, the functions valid for a point nucleus:
\begin{equation}
\begin{split}
g_0 (r)&= 0.00013575124407339550307 r^8\\
&-0.00012633396034194731891 r^7 \\
&+0.0023754193119115541914 r^6\\
&-0.0052460271878852635132 r^5 \\
&+0.16925588925254111005 r^4\\
&-0.25201860708873574898 r^3 \\ 
&+0.95109984162919008905 r^2\\
   &-2.0864972181198001792 r \\
   &+1.6459704071917522632,
   \end{split}
\end{equation}
\begin{equation}
\begin{split}
g_1 (r)&=-0.000078684672329473358699 r^8\\ 
&-0.0012293141869424835524 r^6\\
&-0.097906849416525020713r^4\\
&-0.41666290189975666225 r^2\\
&+0.13769050748433509769
   \end{split}
\end{equation}
and
\begin{equation}
\begin{split}
g_2 (r)&=-0.000012756169252850100497 r^8\\
&+0.017425498169562658160 r^4\\
&+0.44444444460943167625 \,.
   \end{split}
\end{equation}
For $x>3$, we fitted the coefficients in Eq. \eqref{eq:l0asymp} to the numerical values. We obtain 
\begin{align*}
a &= 4.3942926509010\\
b &= -10.059551479890\\
 c &= 5.5493632222582\\ 
 d &= 5.3327556570422\\
 e& = -9.0762837836987\\
 f& = 5.1094977559523 \,.
\end{align*}
Using these functions we reach an agreement to 9 decimal place with both the result of the numerical evaluation and the expansion from \cite{blo1972}.

\section{Coefficients for the numerical evaluation of the Källén and Sabry potential for a finite nucleus.}
\label{app:v21}
The coefficients for the functions defined in Eq. \eqref{eq:L0shape} that we obtained are listed below:
\begin{equation}
\begin{split}
h_0 (r)&= -0.00001608988362060 r^9 \\
& +0.000015791745043 r^8 \\
&-0.00036443366062 r^7 \\
&+0.0008743378646r^6 \\
& -0.038046259798 r^5 \\
&+0.063004651772 r^4 \\
&-0.36332915853 r^3 \\
&+1.04324860906 r^2 \\
&-2.39716878893r+2.005566300 
   \end{split}
\end{equation}
\begin{equation}
\begin{split}
g_1 (r)&=0.7511983817345282548 \\
&+ 0.13888763396658555408 r^2 \\
&+ 0.020975409736870016795 r^4 \\
&+ 0.00017561631242035479319 r^6 \\
&+ 9.057708511987165794\times10^{-6} r^8
   \end{split}
\end{equation}
and
\begin{equation}
\begin{split}
g_2 (r)&=-0.44444444460943167625 \\
& - 0.0034850996339125316319 r^4\\
& - 1.4173521392055667219\times10^{-6} r^8  \,.
   \end{split}
\end{equation}

\end{appendix}

\end{document}